\documentclass[a4paper,11pt]{article}
\pdfoutput=1 
\usepackage{amsmath, amssymb}
\DeclareUnicodeCharacter{2A7E}{\gtrsim}
\usepackage{jcappub} 
\usepackage[T1]{fontenc} 
\usepackage{multirow}
\usepackage{orcidlink}
\usepackage{subcaption}
\usepackage{todonotes}
\usepackage{hyperref}
\usepackage{academicons}
\usepackage{xcolor}
\usepackage{soul}
\usepackage{booktabs}
\sethlcolor{yellow}
\usepackage{orcidlink}
\usepackage{caption}
\usepackage{silence}

\title{Illuminating Dark Energy with Bright Standard Sirens from Future Detectors}

\author[1]{ Samsuzzaman Afroz, \orcidlink{0009-0004-4459-2981}
}
\author[1]{Suvodip Mukherjee, \orcidlink{0000-0002-3373-5236}
}
\author[2,3]{Gianmassimo Tasinato \orcidlink{0000-0002-9835-4864}
}
\affiliation[1]{Department of Astronomy and Astrophysics, Tata Institute of Fundamental Research, Mumbai 400005, India}
\affiliation[2]{Physics Department, Swansea University, SA2 8PP, UK}
\affiliation[3]{Dipartimento di Fisica e Astronomia, Universita di Bologna, INFN, Sezione di Bologna,  viale B. Pichat 6/2, 40127 Bologna, Italy}

\emailAdd{samsuzzaman.afroz@tifr.res.in}
\emailAdd{suvodip@tifr.res.in}
\emailAdd{g.tasinato2208@gmail.com}

\abstract{Understanding the nature and evolution of dark energy (DE) is a central challenge in modern cosmology. In this work, we explore the constraining power of bright standard sirens -- gravitational wave (GW) events with electromagnetic counterparts \text{-} for probing the DE equation of state as function of redshift. Focusing on future GW observations from next-generation ground-based GW detectors such as the Einstein Telescope and Cosmic Explorer, we perform a comprehensive analysis using simulated binary neutron star (BNS) and neutron star-black hole (NSBH) events over five years of observation with a 75\% duty cycle. We consider three broad classes of DE models: (i) phenomenological parametrizations, specifically the Barboza-Alcaniz extension to the Chevallier-Polarski-Linder model; (ii) physically motivated scalar field scenarios, specifically hilltop quintessence; and (iii) evolving dark matter setup in which the matter density evolves as \( (1+z)^{3+\alpha} \). For each case, we jointly infer the Hubble constant \( H_0 \) and model-specific DE parameters from the observed GW luminosity distances and spectroscopic redshifts. Our results demonstrate that bright sirens alone can yield competitive and independent constraints on the time evolution of DE indicating that  
multi-messenger cosmology has the potential to test a wide range of DE theories, bridging phenomenological and physically motivated models, and paving the way for precision cosmology in the era of GW astronomy.}

\begin{document}
\maketitle
\flushbottom

\section{Introduction}

The discovery that the universe is undergoing an accelerated expansion has transformed modern cosmology and introduced one of its deepest unsolved mysteries: the nature of dark energy (DE) \citep{SupernovaSearchTeam:1998fmf,SupernovaCosmologyProject:1998vns}. 
In this work, we explore how future gravitational wave (GW) detectors can help us to shed light on this mystery. 
Accounting for approximately 70\% of the total energy density of the universe, DE is believed to drive this acceleration through a component with negative pressure \citep{Huterer:2017buf,SupernovaCosmologyProject:1996grv,SupernovaSearchTeam:1998bnz}.
While the standard cosmological model, Lambda Cold Dark Matter (\(\Lambda\)CDM) model successfully fits a wide range of observational data, including the cosmic microwave background (CMB), baryon acoustic oscillations (BAO), and Type Ia supernovae, it relies on the simple assumption that DE is a cosmological constant with a constant equation of state \(w = -1\) \citep{Carroll:2000fy,WMAP:2003elm,SDSS:2005xqv,SDSS:2003eyi}. Despite its empirical success, the \(\Lambda\)CDM model raises several unresolved theoretical issues, including the fine-tuning problem (why the vacuum energy density is so small) and the coincidence problem (why the DE density is of the same order of magnitude as the matter density today) \citep{Martin:2012bt,Bernardo:2022cck,Sahni:2002kh,Weinberg:2000yb,Capozziello:2018jya,Abdalla:2022yfr,SDSS:2003eyi}. Emerging evidence for evolving DE \cite{DES:2024jxu,DESI:2024mwx,DESI:2024hhd,DES:2025bxy} raises the compelling possibility that the $\Lambda$CDM paradigm may need to be  reconsidered in the near future.

In fact, theoretical challenges and observational results have led to the exploration of dynamical DE models, in which the equation of state \(w(z)\) varies with redshift. The functional form of \(w(z)\) encodes the underlying physics of the cosmic acceleration and is therefore a critical target for observational cosmology. In this work, we explore how future GW detections can help in determining  \(w(z)\). Phenomenological parametrizations, such as the Chevallier-Polarski-Linder (CPL) model, provide a simple and practical way to study DE dynamics by assuming a two-parameter form for the evolution of \(w(z)\) \cite{Chevallier:2000qy,Linder:2002et,dePutter:2008wt}. While such parametrizations offer ease of comparison with observational data, they are often too restrictive to capture the full complexity of potential DE behavior, especially if the underlying physics involves nonlinear or multi-field dynamics \cite{Wolf:2025jlc}.

To go beyond purely phenomenological descriptions, physically motivated models have been developed in which DE arises from the dynamics of a scalar field evolving in a potential. These models include thawing and freezing quintessence, k-essence, and tracker fields, which can naturally emerge in theories of high-energy physics such as string theory or supergravity \cite{Copeland:2006wr,Zlatev:1998tr,Armendariz-Picon:2000ulo,Brax:1999gp}.

A prominent example is hilltop quintessence, where the scalar field starts near the maximum of a potential and slowly rolls down, resulting in a time-dependent \(w(z)\) that evolves away from \(-1\) as the field thaws. These scenarios provide a more fundamental framework to understand DE, and can connect cosmological behavior with microphysical parameters such as the curvature of the potential and initial field displacement: see e.g. \cite{Dutta:2008qn,Chiba:2009sj,Bhattacharya:2024kxp}.

An even more radical possibility is that the cosmic acceleration is not driven by a new energy component at all, but instead by modifications to the behavior of dark matter or gravity itself. Evolving dark matter models propose that dark matter, usually treated as pressureless, may have a small but dynamically evolving pressure, effectively modifying the background expansion rate. 
This can create effects that resemble those attributed to DE, accommodating recent observational results in a consistent framework \cite{Chen:2025wwn,Kumar:2025etf,Wang:2025zri,Lee:2025hjw}. Similarly, modifications to General Relativity on cosmological scales can alter the relationship between energy density and curvature, leading to accelerated expansion without invoking DE: see e.g. \cite{Joyce:2014kja} for a comprehensive review, and
\cite{Ezquiaga:2018btd} for a review focussed on consequences
for GW observations.

Distinguishing between  competing models of DE requires precise and accurate measurements of the expansion history of the universe across a wide redshift range. Traditionally, this has been pursued using Type Ia supernovae, BAO, and galaxy clustering. However, these methods depend on calibration procedures or assumptions about standard candles and standard rulers, which introduce systematic uncertainties \cite{Afroz:2025iwo, Efstathiou:2024xcq,Efstathiou:2025tie}. In contrast, GW observations offer a new and independent avenue for cosmology \citep{LIGOScientific:2016aoc,LIGOScientific:2017vwq,LIGOScientific:2020iuh,LIGOScientific:2020zkf,LIGOScientific:2019zcs,LIGOScientific:2021aug,LIGOScientific:2017adf,Mukherjee:2024inw}. Compact binary mergers, such as binary neutron star (BNS) or neutron star-black hole (NSBH) systems, emit gravitational radiation that allows for direct measurement of the luminosity distance to the source without the need for intermediate calibration steps \cite{Schutz:1986gp,Holz:2005df,Dalal:2006qt,Arabsalmani:2013bj,Mitra:2020vzq,Vitale:2018wlg,LISACosmologyWorkingGroup:2019mwx,ET:2019dnz,Mukherjee:2020hyn, Diaz:2021pem,Cozzumbo:2024vxw}. (However, these inferred distances can also vary due to the effects of modified gravity theories \cite{Belgacem:2018lbp,LISACosmologyWorkingGroup:2019mwx,Mukherjee:2020mha,Afroz:2024joi,Afroz:2024oui,Afroz:2023ndy}.) When an electromagnetic (EM) counterpart is identified, the host galaxy redshift can be measured spectroscopically, turning the event into a bright standard siren.

The use of bright standard sirens opens up the possibility of mapping the distance-redshift relation with unprecedented accuracy and in a model-independent fashion. This provides a unique opportunity to probe the DE equation of state and distinguish between different theoretical scenarios. Future ground-based GW detectors, such as the Einstein Telescope (ET) \cite{Punturo:2010zz,Branchesi:2023mws,Abac:2025saz} and Cosmic Explorer (CE) \cite{Reitze:2019iox,Evans:2021gyd}, are expected to detect thousands of BNS and NSBH mergers out to redshifts \(z \sim 3\)-\(4\), many of which are expected to have observable EM counterparts. The spectroscopic identification of these hosts enables a precise reconstruction of the Hubble parameter and the evolution of \(w(z)\) with cosmic time: see e.g. \cite{Chen:2024gdn} for a review. In this respect, future generation of GW detectors after 2040 will then be able to do better than LVK or LISA, which will nevertheless able to measure the Hubble parameter
at percent level in a range of redshifts \cite{Mangiagli:2023ize,Salvarese:2025qel}.

In this work, we focus exclusively on bright sirens-GW events with confidently identified EM counterparts-and explore their utility in constraining the nature of DE from the next-generation (XG) detectors. We adopt a model-driven approach and consider three broad classes of DE scenarios: (i) phenomenological parametrizations, focusing specifically on the Barboza--Alcaniz model; (ii) physically motivated scalar field models, with emphasis on hilltop quintessence; and (iii) evolving dark matter models, in which a deviation from the standard matter scaling law introduces effective DE-like behavior. For each of these scenarios, we use a hierarchical bayesian inference framework to jointly constrain the Hubble constant and the model-specific parameters using simulated GW data from BNS and NSBH populations. Our analysis incorporates realistic luminosity distance uncertainties, including both instrumental noise and weak lensing effects, and assumes spectroscopic redshifts with negligible uncertainty.

By comparing the inferred posterior distributions and the reconstructed evolution of \(w(z)\), we evaluate the ability of bright sirens to discriminate between these different models. Our results show that even with bright sirens alone the time evolution of DE can be constrained at high precision. This highlights the potential of next-generation GW observatories as powerful cosmological probes and emphasizes the importance of EM follow-up in maximizing the scientific return from GW events. 

This paper is organized as follows. In Section~\ref{sec:theory}, we outline the theoretical background and present the three classes of dark energy models. Section~\ref{sec:MockSamples} describes the simulation of GW events. Section~\ref{sec:results} presents the main results for each model class. We conclude in Section~\ref{sec:conclusion} with a discussion of the implications of our findings for future GW cosmology.

\section{Dark Energy: Theory and Classification}
\label{sec:theory}

The accelerated expansion of the universe, first revealed by observations of Type Ia supernovae and further confirmed by CMB measurements and large scale structure surveys, strongly suggests the existence of a component with negative pressure commonly termed DE. Despite two decades of progress, the fundamental nature of DE remains one of the most profound and unresolved questions in modern cosmology. We now turn to examining promising and well-motivated scenarios for evolving DE. In the following sections, we explore how these models can be tested using future GW observations.

Within the standard Friedmann-Lemaitre-Robertson-Walker (FLRW) framework, the evolution of the universe is described by the Friedmann equations,
\begin{equation}
H^2(t) = \left(\frac{\dot{a}}{a}\right)^2 = \frac{8\pi G}{3} \rho - \frac{k}{a^2},
\end{equation}
where $H(t)$ is the Hubble parameter, $a(t)$ is the scale factor, $\rho$ denotes the total energy density, and $k$ represents the spatial curvature. Under the well supported assumption of a spatially flat universe $k=0$ (but see
e.g. \cite{Bhattacharya:2024hep} for a recent work analysing recent data without relying on this assumption), the Friedmann equation simplifies to
\begin{equation}
H^2(z) = H_0^2 \left[ \Omega_{m,0}(1+z)^3 + \Omega_{r,0}(1+z)^4 + \Omega_{\text{DE}}\, f(z) \right],
\end{equation}
where $H_0$ is the present day Hubble constant, and $\Omega_{m,0}$, $\Omega_{r,0}$, and $\Omega_{\text{DE}}$ are the present day density parameters for matter, radiation, and DE, respectively. The function $f(z)$ captures the redshift evolution of the DE density and is defined by its equation of state parameter $w(z)$,
\begin{equation}
f(z) = \exp \left[3 \int_0^z \frac{1+w(z')}{1+z'} dz'\right],
\end{equation}
where $w(z) = p_{\text{DE}} / \rho_{\text{DE}}$. At early times, radiation ($\rho_r \propto a^{4}$) and matter ($\rho_m \propto a^{3}$) dominate the cosmic energy budget and drive the expansion; DE becomes dynamically important only at late times ($z \lesssim 1$), when its negative pressure begins to accelerate the universe.

The impact of DE on observable quantities is profound, influencing both the background expansion and the growth of structure. For example, the luminosity distance to a source at redshift $z$ is given by
\begin{equation}
D_L(z) = (1+z) \int_0^z \frac{c}{H(z')} dz',
\end{equation}

linking observations directly to the expansion history determined by $w(z)$. While a cosmological constant ($w = -1$) remains consistent with current observations, the possibility of a time-varying equation of state -- arising from more complex underlying physics -- cannot be excluded; indeed, it may even be favored by recent cosmological data \cite{DES:2024jxu,DESI:2024mwx,DESI:2024hhd,DES:2025bxy}.

Theoretically, a wide variety of models have been proposed to account for DE. In many approaches, DE is treated as a perfect fluid with an equation of state that evolves according to the conservation equation,
\begin{equation}
\dot{\rho}_{\text{DE}} + 3H(1+w)\rho_{\text{DE}} = 0,
\end{equation}
which yields the redshift evolution described above. However, the underlying physics that gives rise to this evolution may vary dramatically across different models. These range from phenomenological parametrization of $w(z)$ which are designed for flexibility and efficient data fitting to physically motivated scenarios involving scalar fields or modifications to gravity, and even to unified dark sector frameworks where dark matter itself may have nontrivial dynamics.

For instance, scalar field theories  such as thawing and freezing quintessence, k-essence, and tracker models provide natural candidates for dynamical DE and can emerge from high energy theory. At the same time, models such as the Chaplygin gas and those involving evolving dark matter illustrate that the observed acceleration could, in principle, arise from exotic properties or interactions within the dark sector itself -- rather than requiring the introduction of a new energy component. See e.g. \cite{Joyce:2014kja}
for a comprehensive  review.

Given this broad theoretical landscape, it is essential to systematically explore different classes of models phenomenological parametrization, physically motivated scalar field models, and evolving dark matter scenarios in order to robustly interpret cosmological data and gain insights into the nature of cosmic acceleration. In the following subsections, we discuss each of the three approaches in detail, establishing the foundation for the models explored in this work, which are testable by GW experiments.

\subsection{Phenomenological Parametrization}
\label{sec:thphpar}

Phenomenological parametrization offer a practical and model independent way to describe the possible redshift evolution of the DE equation of state parameter, $w(z)$. Instead of relying on an underlying physical theory, these models are constructed to efficiently fit cosmological data and to capture a wide variety of DE behaviors with a small set of parameters. See
e.g. \cite{Wolf:2025jlc} for a recent nice account. One of the most widely used forms is the Chevallier-Polarski-Linder (CPL) parametrization,

\begin{equation}
    w(z) = w_0 + w_a \frac{z}{1+z},
\end{equation}
where $w_0$ denotes the present day value of the equation of state and $w_a$ describes its evolution with redshift. This form, which is linear in the scale factor $a = 1/(1+z)$, provides a simple yet effective way to interpolate between early and late time behavior of DE.

An alternative is the Jassal-Bagla-Padmanabhan (JBP) parametrization \cite{Jassal:2005qc},
\begin{equation}
    w(z) = w_0 + w_a \frac{z}{(1+z)^2},
\end{equation}
which ensures that $w(z)$ returns to $w_0$ both at $z \to 0$ and $z \to \infty$, and exhibits a slower evolution at intermediate redshifts compared to CPL. For improved high redshift behavior, the Barboza--Alcaniz (BA) parametrization is often employed \cite{Barboza:2008rh}
\begin{equation}
    w(z) = w_0 + w_a \frac{z(1+z)}{1+z^2},
    \label{eq:BAParametrization}
\end{equation}
which remains bounded and smoothly connects the present and the past, making it particularly suitable for probing the equation of state over an extended range of redshifts.

To introduce greater flexibility and better account for potential nonlinear evolution of DE, an extended version of the CPL parametrization has been proposed in \citep{Afroz:2024lou}, expressed as:
\begin{equation}
    w(a) = w_0 + w_a\biggl(\frac{z}{1+z}\biggr) + w_b\biggl(\frac{z}{1+z}\biggr)^2,
    \label{eq:ExtCPL}
\end{equation}
where \( w_0 \) denotes the present-day value of the EoS, while \( w_a \) and \( w_b \) capture its linear and quadratic evolution as a function of redshift. This three-parameter extension improves the ability to model DE behavior across both low and high redshift regimes and allows for possible departures from simpler two-parameter descriptions.

\subsection{Physically Motivated Scalar Field Models}
\label{sec:phymot}

Physically motivated models of DE frequently invoke a dynamical scalar field, $\phi$, minimally coupled to gravity and evolving in a potential $V(\phi)$. The evolution of the field is governed by the Klein Gordon equation,
\begin{equation} 
    \ddot{\phi} + 3H\dot{\phi} + \frac{dV}{d\phi} = 0,
\end{equation}
where $H$ is the Hubble parameter. The corresponding energy density and pressure are
\begin{align}
    \rho_\phi &= \frac{1}{2}\dot{\phi}^2 + V(\phi), \\
    p_\phi &= \frac{1}{2}\dot{\phi}^2-  V(\phi),
\end{align}
leading to the equation of state parameter,
\begin{equation}
    w_\phi = \frac{p_\phi}{\rho_\phi} = \frac{\dot{\phi}^2/2  -V(\phi)}{\dot{\phi}^2/2 + V(\phi)}.
\end{equation}

Depending on the form of the potential, a rich variety of cosmological behaviors can emerge: see e.g. the review \cite{Copeland:2006wr}. For instance, in thawing quintessence models, the scalar field remains nearly frozen by Hubble friction near the top of a flat or hilltop potential for much of cosmic history, with $w_\phi \approx -1$ at early times \cite{Wetterich:1987fm,Ratra:1987rm, Caldwell:1997ii}. As the universe expands and the Hubble parameter decreases, the field eventually “thaws” and rolls down the potential, resulting in a time dependent equation of state that deviates from the cosmological constant value. In contrast, freezing quintessence models feature a field that rolls more rapidly at early times, with $w_\phi > 1$, but slows and approaches $w_\phi \rightarrow 1$ at late times: see e.g. \cite{Garcia-Garcia:2019cvr,Wolf:2024eph,Tada:2024znt,Wolf:2023uno,Shlivko:2024llw}. Tracker and attractor models can interpolate between these regimes and help alleviate sensitivity to initial conditions. More generally, other frameworks such as k-essence, coupled quintessence, and non canonical kinetic terms have been explored for their potential to explain cosmic acceleration with additional degrees of freedom.

In this work, among the many possibilities, we focus on the hilltop quintessence model, a well motivated example of the thawing class, which fits well with quantum gravity and effective-field theory considerations \cite{Bhattacharya:2024kxp,Borghetto:2025jrk}. In the hilltop scenario, the scalar field is initially located near the maximum of its potential and evolves slowly due to the flatness of the potential and the influence of Hubble friction in the early universe. This setting is theoretically appealing, as it arises naturally in axion like models and effective field theory constructions, and can be compatible with certain quantum gravity constraints. The field’s slow roll away from the hilltop ensures that the equation of state remains close to one at early times, then increases as the field thaws. The analytic expression for the equation of state parameter is given by \cite{Dutta:2008qn}
\begin{equation}
1 + w(a) = (1 + w_0)a^{3(K-1)} 
\frac{\left[(F(a)+1)^K(K-F(a)) + (F(a)-1)^K(K+F(a))\right]^2}
     {\left[(\Omega_{\phi0}^{-1/2}+1)^K(K-\Omega_{\phi0}^{-1/2}) + (\Omega_{\phi0}^{-1/2}-1)^K(K+\Omega_{\phi0}^{-1/2})\right]^2},
\end{equation}
where
\begin{align}
F(a) &= \sqrt{1 + (\Omega_{\phi0}^{1}1)a^{3}},\nonumber \\
K &= \sqrt{1  \frac{4}{3} \frac{V''(\phi^*)}{V(\phi^*)}},
\end{align}
with $\Omega_{\phi0}$ denoting the present day scalar field energy density, $w_0$ the present day value of the equation of state, and $V''(\phi^*)$ the second derivative of the potential at its maximum. 

We choose the hilltop quintessence model because it is physically well motivated from the perspective of high energy theory, naturally yields the thawing dynamics of DE, and allows for analytic expressions that are straightforward to compare with observations. Moreover, it serves as a bridge between micro-physical theory and cosmological data by relating the behavior of $w(a)$ to the curvature of the potential, $V''/V$, at the hilltop. This makes the hilltop quintessence scenario not only theoretically appealing but also practically useful as a benchmark for studying departures from the cosmological constant in a controlled and predictive way. Nevertheless, we emphasize that the methods developed here are sufficiently general to test a broad class of dark energy models with gravitational wave experiments.

\begin{figure}[ht]
    \centering
    \includegraphics[width=11.0cm, height=7.0cm]{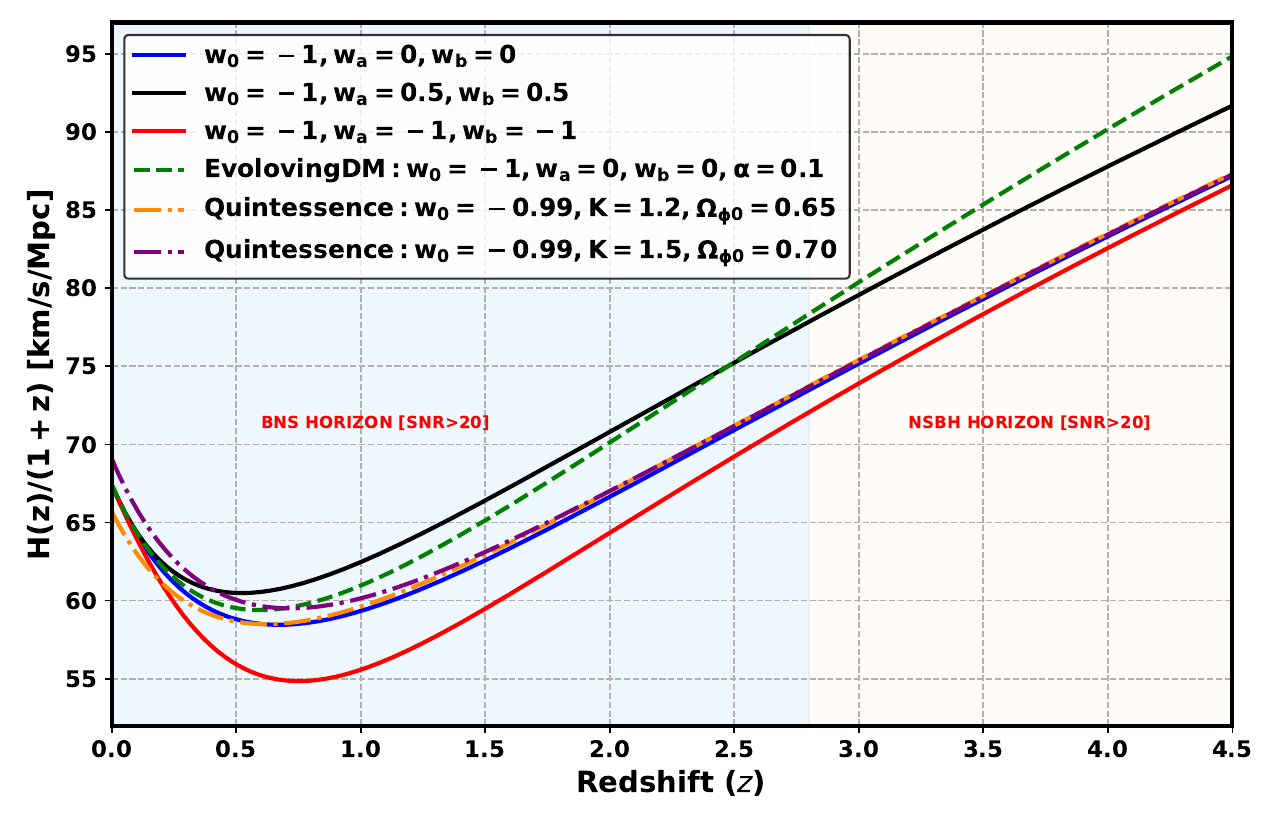}
    \caption{\small Theoretical evolution of the scaled Hubble parameter, \( H(z)/(1+z) \), as a function of redshift for various DE models and parameter choices. The curves illustrate different scenarios: CPL-like extensions with increasing \( w_0, w_a, w_b \), an evolving dark matter model with \( \alpha = 0.1 \), and two hilltop quintessence cases with different curvature \( K \) and present-day scalar field energy densities \( \Omega_{\phi0} \). The shaded regions indicate the approximate detection horizons (SNR $>$ 20) for BNS and NSBH systems for next-generation GW detectors. This plot provides a comparative theoretical expectation for how modified DE dynamics affect the expansion history, and highlights the redshift reach of GW sources used in our analysis.}
\end{figure}

\subsection{Evolving Dark Matter Models}
\label{sec:evdm}

A compelling alternative to modifications in the DE sector is the hypothesis that dark matter itself may possess nontrivial dynamics, characterized by a time dependent equation of state, $w_{DM}(z)$. In this framework, instead of assuming that all dark matter behaves as perfectly cold and pressure less ($w_{DM} = 0$), a sub component or the entirety of dark matter is allowed to have a small, but potentially evolving, pressure. This pressure can remain negligible in the early universe --  thereby preserving the standard successes of cold dark matter in explaining the formation of large scale structure and the features of the cosmic microwave background -- but may become significant at late times as $w_{DM}(z)$ deviates from zero. 

The motivation for considering evolving dark matter stems from both theoretical and observational directions. On the theoretical side, the nature of the dark sector is still unknown, and it is plausible that new physics such as dark sector interactions, scalar fields, or composite fluids could naturally lead to a small and evolving pressure. Observationally, some features in cosmological data that are typically interpreted as evidence for a dynamical or even phantom like DE component could, in principle, be reproduced by allowing dark matter to evolve away from the standard $w_{DM}=0$ case: see e.g.  \cite{Chen:2025wwn,Kumar:2025etf,Wang:2025zri,Lee:2025hjw}. 

Within this formalism, the contribution of dark matter to the cosmic expansion changes accordingly, and we express it  as $\Omega_{m,0}(1+z)^{3+\alpha}$, where the parameter $\alpha$ encodes the effective deviation from standard scaling. For $\alpha = 0$, one recovers the usual cold dark matter evolution; for nonzero $\alpha$, the effective behavior of dark matter can enhance or dilute its density at late times, thereby mimicking the effect of DE or contributing to the accelerated expansion. This approach provides a unified framework in which the accelerated expansion arises not solely from an exotic DE fluid but as a possible manifestation of richer dynamics within the dark matter sector itself.

\smallskip

The scenarios discussed thus far in this section can be tested and constrained through a combination of current and forthcoming cosmological observations. In the following sections, we examine the role that future GW measurements will play in this endeavour.

\section{Gravitational Waves  mock samples}
\label{sec:MockSamples}
Gravitational wave observations from standard sirens \cite{Schutz:1986gp,Holz:2005df,Dalal:2006qt,Arabsalmani:2013bj,Mitra:2020vzq,Vitale:2018wlg,LISACosmologyWorkingGroup:2019mwx,ET:2019dnz,Mukherjee:2020hyn,Diaz:2021pem,Cozzumbo:2024vxw} offer the potential to measure the Hubble constant with percent-level precision \cite{Chen:2017rfc}. In particular, detections by LVK and LISA over the next decade or two are expected to play a central role in this effort \cite{LISA:2024hlh}; see, for example, \cite{Afroz:2024oui, Salvarese:2025qel} for a recent analysis.  The physics of DE, nevertheless,
is set to be revolutionized by next-generation ground-based detectors such as the CE\cite{Reitze:2019iox,Evans:2021gyd} and the ET \cite{Punturo:2010zz,Branchesi:2023mws,Abac:2025saz}. These detectors are designed to vastly expand the observable volume of the universe and probe the low-frequency regime critical for precision cosmology. The CE, slated for construction in the United States, will have arm lengths of either 20 or 40 kilometers, offering a significant boost in sensitivity to GW signals from sources at high redshift. This improvement will enable more accurate estimation of source parameters. In parallel, the ET, a European initiative featuring underground interferometer arms, is designed to suppress seismic noise and extend observational capability across a wide frequency band—from a few Hz up to several kHz. These upcoming detectors will not only deepen our insights into compact binary mergers but will also create unique prospects for probing the nature of DE through standard siren observations.

Reconstructing the EoS of DE demands a carefully modeled and statistically representative mock GW catalog. The success of such a reconstruction hinges on an accurate characterization of the underlying astrophysical GW source population. The number of observed GW events will be influenced by critical factors like merger rates, redshift evolution, and the mass distribution of binaries. Future observatories such as CE and ET, with their improved reach and sensitivity, will be capable of detecting mergers across a broader redshift interval. This positions them as powerful tools for cosmological inference that operate independently of traditional electromagnetic probes.

We perform the DE EoS reconstruction using bright sirens. Bright sirens correspond to GW events with EM counterparts, such as BNS and NSBH mergers, where the redshift can be directly measured from their EM counterpart observations. A precise understanding of the population properties of GW sources is essential, as these properties directly affect the inferred redshift range and the accuracy of the reconstructed Hubble parameter ${H}(z)$ and the DE EoS, $w(z)$. Our analysis incorporates detailed astrophysical models that capture GW source distributions, merger rates, and potential selection effects. These models are specifically tailored to the anticipated capabilities of next-generation GW observatories like CE and ET. In what follows we  systematically discuss the key ingredients of these GW mock samples, their statistical properties, and how they influence the reconstruction analysis. Our approach ensures the robustness of our cosmological analysis while accounting for the physical and observational factors that underpin GW detections.

\texttt{Mass model:} The mass distribution model for GW sources, including NSBH and BNS systems, is guided by recent insights from the third GW catalog released by the LVK collaboration~\cite{talbot2018measuring, abbott2019binary}. For black holes, we utilize the \textbf{Power Law + Gaussian Peak} prescription~\cite{abbott2023population}, which combines a power-law component for the primary mass spectrum with a Gaussian peak capturing the excess population of black holes in the \({5M_{\odot}} \) to \({20M_{\odot}} \) range. To ensure a smooth and realistic distribution, the model also incorporates additional smoothing functions aligned with current observational trends. In the case of neutron stars, we adopt a uniform distribution spanning \({1M_{\odot}} \) to \({2M_{\odot}} \), consistent with the relatively constrained mass range observed in BNS systems. The integration of these distributions yields a robust and representative framework for modeling the mass properties of NSBH and BNS populations used in our study.

\texttt{Merger rate:} To model the merger rates of compact binary systems, including BNS and NSBH, we employ a delay time distribution framework \cite{o2010binary, dominik2015double}. This model characterizes the merger rate as a function of the time delay between the formation of the progenitor binary system and its eventual coalescence. The delay time depends on the stellar evolution and dynamical processes of the system and is not uniform across all binaries. Different classes of compact binaries can follow distinct evolutionary tracks, resulting in varied delay time distributions. For example, BNS and NSBH systems often undergo different formation and evolutionary histories compared to BBH systems, due to differences in supernova mechanisms, stellar structure, and environmental influences. These distinctions translate into differing merger rates and distributions of delay times across binary types. Accurate modeling of these distributions is crucial for interpreting the population of GW events and understanding their astrophysical origins. The delay time distribution functions adopted in this work are designed to reflect these variations, providing a statistically robust description across all considered compact binary systems. The functional form of the distribution accounts for the range and nature of delay times and is given by:

\begin{equation}
    p_t(t_d|t_d^{min},t_d^{max},d) \propto 
    \begin{cases}
    (t_d)^{-d} & \text{ for }  t_d^{min}<t_d<t_d^{max}, \\
    0 & \text{otherwise}.    
    \end{cases}
    \label{eq_ptdel}
\end{equation}
The delay time is given by ${t_d=t_m-t_f}$, where ${t_m}$ and ${t_f}$ are the look-back times of merger and formation respectively \cite{karathanasis2023binary}. Hence,  the merger rate at redshift ${z}$ can be defined as 
\begin{equation}
    R_{TD}(z)=R_0\frac{\int_z^{\infty}p_t(t_d|t_d^{min},t_d^{max},d)R_{SFR}(z_f)\frac{dt}{dz_f}dz_f}{\int_0^{\infty}p_t(t_d|t_d^{min},t_d^{max},d)R_{SFR}(z_f)\frac{dt}{dz_f}dz_f}. 
\end{equation}
The parameter \( {R_0} \) represents the local merger rate, which quantifies the frequency of mergers occurring at a redshift of \( {z = 0} \). According to the study
\cite{abbott2023population},
the estimated \( {R_0} \) values for BNS systems varies between 10 \( {Gpc^{-3}\,yr^{-1}} \) and 1700 \( {Gpc^{-3}\,yr^{-1}} \).
In the case of neutron star–black hole (NSBH) systems, the \( {R_0} \) values are estimated to lie between 7.8 \( {Gpc^{-3}\,yr^{-1}} \) and 140 \( {Gpc^{-3}\,yr^{-1}} \). 

For the purposes of our study, we adopt a standard local merger rate of \( {R_0} = 20\,{Gpc^{-3}\,yr^{-1}} \) uniformly for BNS, NSBH, and BBH systems. The numerator of the expression involves the integration over redshift ${z_f}$ from ${z}$ to infinity, where $${p_t(t_d|t_d^{min},t_d^{max},d)},$$ is the delay time distribution, and ${\frac{dt}{dz_f}}$ is the Jacobian of the transformation. The star formation rate  denoted by ${R_{SFR}(z)}$ is determined using the Madau Dickinson model \cite{madau2014cosmic}. 
\begin{figure}[ht]
    \centering
    \includegraphics[width=11.0cm, height=7.0cm]{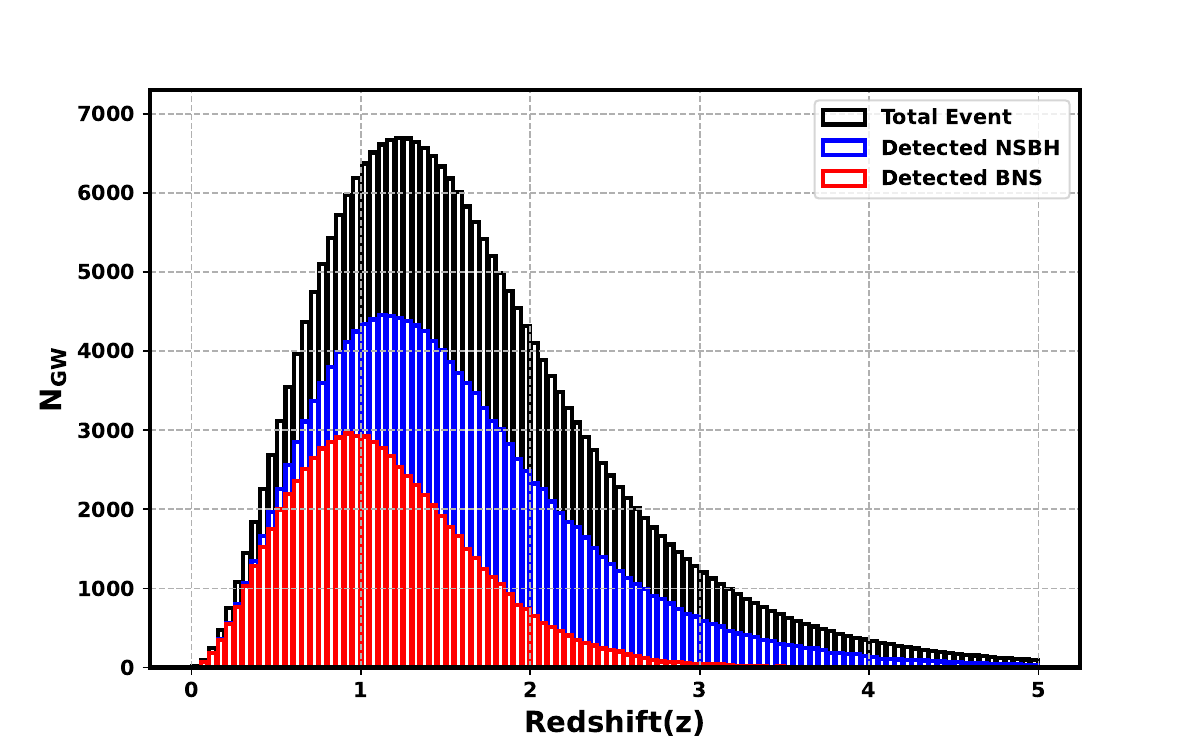}
    \caption{\small The figure illustrates the total number of simulated merger events (in gray) alongside the detectable events for BNS (in black), and NSBH (in blue) using the CE+ET detector network. The analysis assumes a five-year observation period with a 75\% duty cycle. The total number of simulated merger events is 244,832, of which 152,054 NSBH, and 75,832 BNS events are detected.}
    \label{fig:EventCount}
\end{figure}
The total number of compact binary coalescing events per unit redshift is estimated as 
\begin{equation}
\frac{dN_{GW}}{dz} = \frac{R_{TD}(z)}{1+z} \frac{dV_c}{dz}(z) T_{obs},
\label{eq:totno}
\end{equation}
where $T_{obs}$ indicates the total observation time, ${\frac{dV_c}{dz}}$ corresponds to the differential comoving volume element, and ${R(z)}$ denotes the merger rate. We consider the delay time merger rate with a specific minimum delay time ${t_d =500\,\text{Myrs}}$ and a power-law exponent of ${d=1}$ 
in Equation \eqref{eq_ptdel} \cite{karathanasis2022gwsim}. To determine which events are detectable, the calculation of the matched filtering signal-to-noise ratio (SNR) plays a crucial role. The SNR serves as a measure of the strength of the GW signal relative to the background noise. Only those events with a matched filtering SNR greater than or equal to a predetermined threshold SNR (${\rho_{TH}}$) can be reliably detected \cite{maggiore2007gravitational}.

For GW signal emitted by an optimally oriented binary system, the optimized SNR  ${\rho}$, is defined as  
\begin{equation}
\rho^2 \equiv 4\int_{f_{\text{min}}}^{f_{\text{max}}} df \frac{|h(f)|^2}{S_n(f)},
\label{snr}
\end{equation}
here, ${S_n(f)}$ represents the noise power spectral density of the detector \cite{sathyaprakash1991choice, cutler1994gravitational, balasubramanian1996gravitational}. The function ${h(f)}$ corresponds to the GW strain in the restricted post-Newtonian approximation and is defined for plus (${+}$) and cross (${\times}$) polarization as \cite{Maggiore:2007ulw,ajith2008template}

\begin{equation}
    h(f)_{\{+, \times\}}=\sqrt{\frac{5\eta}{24}}\frac{(GM_c)^{5/6}}{D_L\pi^{2/3}c^{3/2}}f^{-7/6}e^{\iota\Psi(f)}\mathcal{I}_{\{+, \times\}}.
\end{equation}

In this expression, the symbol $\eta$ represents the symmetric mass ratio. The term $M_c$ is  the chirp mass of the system. The variable $D_L$ denotes the luminosity distance. The constant c represents the speed of light in a vacuum. $\mathcal{I}_{+}= (1+\cos^{2}i)/2$ and $\mathcal{I}_{\times}= \cos i$ depends on the inclination angle $i$. Finally, $\Psi(f)$ stands for the phase of the waveform. However, the signal detected by a GW detector $h_{det}$ is a complex interplay of several variables, including the detection antenna functions ($F_{+}, F_{\times}$), and can be expressed as \cite{finn1993observing} 
\begin{equation}
    h_{det}=F_{+}h_{+}+F_{\times}h_{\times},
\end{equation}
here $F_{+}$ and $F_{\times}$ are the antenna functions defined as follows
\begin{align}\label{eq:antenna}
    &F_{+}\,=\,\frac{1}{2}(1+cos^2\theta)\,cos2\phi\, cos2\psi-cos\theta\, sin2\phi\, sin2\psi,\\\nonumber
     &F_{\times}\,=\,\frac{1}{2}(1+\cos^2\theta)\,\cos2\phi\, sin2\psi+cos\theta \,sin2\phi\, cos2\psi,
\end{align}
where $\theta$ and $\phi$ define the location of the source in the sky, and ${\psi}$ is related to the orientation of the binary system with respect to the detector. Consequently, the matched filtering SNR $\rho$ takes the form 

\begin{equation}
    \rho = \frac{\Theta}{4}\biggl[4\int_{f_{min}}^{f_{max}}h(f)^2/S_n(f)df\biggr]^{1/2},
    \label{eq:SNR}
\end{equation}
where $\Theta^2 \equiv 4 \left(F_{+}^2(1+\cos^2i)^2 + 4F_{\times}^2\cos^2i\right)$. Averaging over many binaries inclination angle and sky positions, \cite{finn1996binary} shows that $\Theta$ follows a distribution

\begin{equation}
P_{\Theta}(\Theta) = 
\begin{cases}
5\Theta(4-\Theta)^3/256& \text{if } 0<\Theta<4,\\
0,              & \text{otherwise}.
\end{cases}
\end{equation}

We consider the next-generation GW detector network formed by the combination of CE and the ET. Specifically, we focus on a configuration featuring a 40\,km arm-length CE along with a triangular ET layout, which we collectively refer to as the CEET setup. A SNR threshold of \( \rho_{\text{Th}} = 20 \) is employed. Event simulations are performed using source distances inferred from redshift values, the angular parameter \( \Theta \), and binary component mass distributions. A redshift binning of size \( \Delta z = 0.05 \) is adopted, within which the total number of mergers is calculated using Equation~\eqref{eq:totno}. Assuming a local merger rate of \( R_0 = 20\,\mathrm{Gpc}^{-3}\,\mathrm{yr}^{-1} \), an observation time of five years, and a duty cycle of 75\%, the CEET network is expected to detect a substantial number of events over a broad redshift range. Figure~\ref{fig:EventCount} displays the predicted total and detectable numbers of NSBH and BNS events across redshift bins of width \( \Delta z = 0.05 \). To compute these estimates, the number of mergers in each redshift interval is obtained via Equation~\eqref{eq:totno}. For each event, binary component masses and the parameter \( \Theta \) are drawn from their respective probability distributions using the inverse transform sampling method. Black hole masses are sampled in the range \( 5M_{\odot} \) to \( 20M_{\odot} \), while neutron star masses are drawn uniformly between \( 1M_{\odot} \) and \( 2M_{\odot} \). The angular parameter \( \Theta \) is sampled from the interval \([0, 4]\). Given the redshift of each event, the corresponding luminosity distance is calculated, and the single-detector SNR is obtained via Equation~\eqref{eq:SNR}. The combined network SNR, \( \rho_{\text{total}} \), is then calculated using the standard relation \( \rho_{\text{total}} = \sqrt{\sum_i \rho_i^2} \), summing over individual detectors. This procedure reflects the detection sensitivity of the CEET configuration over a five-year period with a 75\% duty cycle. The resulting event distribution, shown in Figure~\ref{fig:EventCount}, offers insight into the expected temporal and redshift-dependent occurrence of compact binary mergers and their detectability with upcoming GW observatories.

\texttt{Parameter Estimation using Bilby:} We begin the parameter estimation for the selected GW sources using the \texttt{Bilby} package \cite{ashton2019bilby}, which yields realistic posterior distributions for the GW luminosity distance, marginalized over the remaining source parameters. The distribution of source masses and the number of events across redshift are assigned following the methods outlined earlier. For other parameters such as the inclination angle ($i$), polarization angle ($\psi$), GW phase ($\phi$), right ascension (RA), and declination (Dec), we assume uniform priors, under the assumption of non-spinning systems. GW signals are simulated using the \texttt{IMRPhenomHM} waveform model \cite{kalaghatgi2020parameter}, which incorporates higher-order modes to mitigate the degeneracy between luminosity distance ($D_L$) and inclination angle ($i$). We fix the priors for all parameters--except for $m_1$, $m_2$, $D_L$, $i$, RA, and Dec--as delta functions, allowing us to extract detailed posterior distributions. Among these, the posterior on $D_L$ plays a crucial role in our study, as it directly impacts the reconstruction of dark energy properties.

\section{Results}
\label{sec:results}
We now present the results of our inference for DE models using simulated GW observations from standard sirens. We focus on the three classes of scenarios
discussed in Section~\ref{sec:theory}: (1) phenomenological parametrizations, (2) physically motivated scalar field models, and (3) evolving dark matter scenarios. In each case, we perform a joint inference of the DE EoS parameters along with the Hubble constant \( H_0 \), within a hierarchical bayesian framework. The posterior distribution is constructed using the observed GW luminosity distances \( D_L\), their associated uncertainties, and redshift information obtained from EM counterparts (i.e., bright sirens). For bright sirens, spectroscopic redshifts are assumed, allowing the redshift posterior to be modeled as a delta function. The fiducial model in all cases is taken to be \( \Lambda \)CDM, with cosmological parameters fixed to the \textit{Planck} 2015 best-fit values~\cite{Planck:2015fie}.

\subsection{Phenomenological Parametrizations}
\label{sec:PhenomenologicalResults}

We begin with the Barboza-Alcaniz (BA) parametrization of the DE EoS, as defined in Equation~\ref{eq:BAParametrization}. This parameterization ensures that \( w(z) \) remains well-behaved across the entire redshift range. Within the Bayesian framework, we infer the parameters \( H_0, w_0, w_a \) by evaluating the joint posterior:
\begin{equation}
\begin{aligned}
    P(H_0, w_0, w_a) &\propto \Pi(H_0) \Pi(w_0) \Pi(w_a) \prod_{i=1}^{n_{GW}} \mathcal{L} \left(D_L^{i} \mid H_0, w_0, w_a, z^i \right),
\end{aligned}
\label{eq:BAposterior}
\end{equation}

where \( \mathcal{L}(D_L^i \mid H_0, w_0, w_a, z^i) \) represents the likelihood for each GW event, and \( n_{\text{GW}} \) is the total number of sources in the catalog. The terms \( \Pi(H_0), \Pi(w_0), \Pi(w_a) \) denote the prior distributions for the cosmological parameters, for which we assume flat priors throughout this analysis. Given the high precision of spectroscopic redshift measurements, we neglect marginalization over redshift in this analysis.

\begin{figure}[ht]
    \centering
    \includegraphics[width=10.0cm, height=9.0cm]{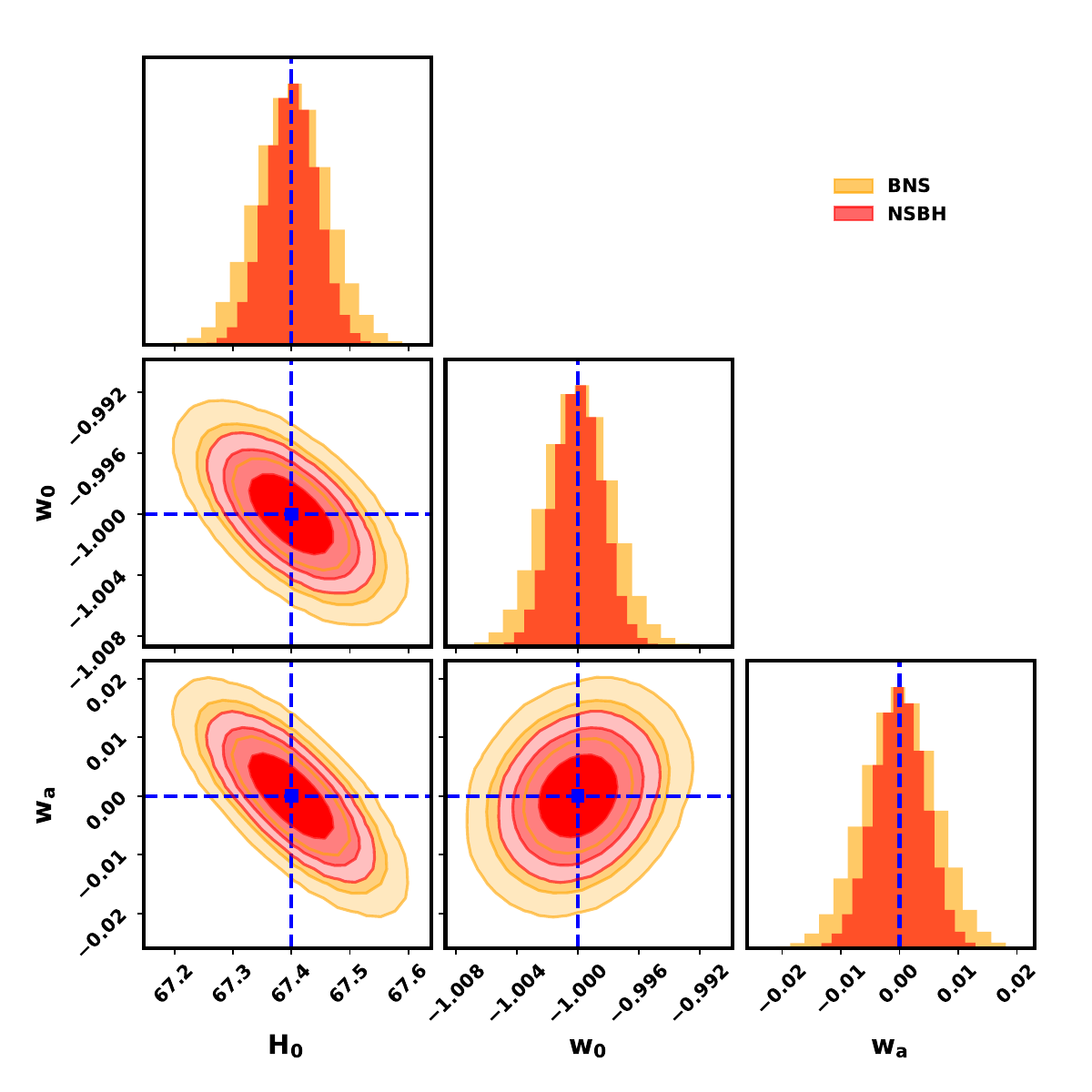}
    \caption{\small Joint posterior distributions of the Barboza--Alcaniz parameters \( w_0 \), \( w_a \), and \( H_0 \) inferred from simulated bright siren observations involving BNS and NSBH sources. The contours correspond to the 68.27\% (1$\sigma$), 95.45\% (2$\sigma$), and 99.73\% (3$\sigma$) confidence levels for the two-dimensional marginalized posteriors.}
    \label{fig:BACornerPlot}
\end{figure}

\begin{figure}[ht]
    \centering
    \includegraphics[width=11.0cm, height=10.0cm]{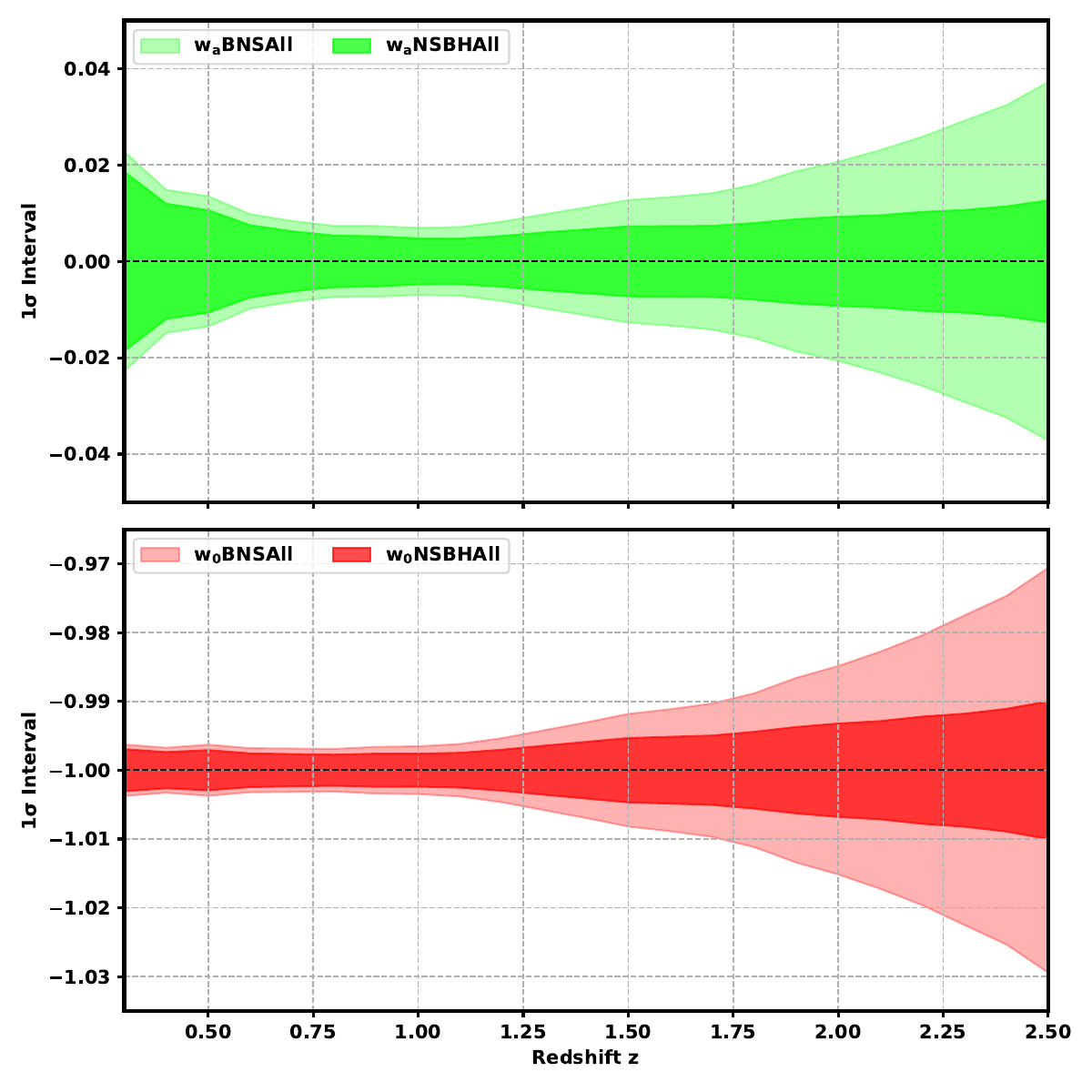}
    \caption{\small Reconstructed evolution of the DE equation of state \( w(z) \) in the BA model, using bright siren data. Shaded bands indicate the 1\(\sigma\) credible regions for BNS and NSBH populations.}
    \label{fig:BAParamEvo}
\end{figure}

The dominant source of uncertainty in constraining the EoS parameters arises from the measurement error in \( D_L\). The precision of GW-based distance measurements depends on detector sensitivity and network configuration, and is further degraded by weak lensing especially for high-redshift sources observable by future detectors like CE and ET. Gravitational lensing magnifies or demagnifies the signal, modifying the observed luminosity distance as \cite{Mpetha:2022xqo}
\[
D_L^{obs} = \frac{D_L^{true}}{\sqrt{\mu}},
\]
where \( \mu \) is the lensing magnification. This introduces a systematic bias in the observed distance:
\[
\frac{\Delta D_L}{D_L} = 1 - \frac{1}{\sqrt{\mu}}.
\]
To incorporate this effect, we convolve the anticipated lensing-induced uncertainty with the magnification probability distribution \( p(\mu) \), derived under a standard $\Lambda$CDM cosmology, significantly influences the overall measurement accuracy\footnote{It our important to note that DE model beyond cosmological constant can lead to a modified uncertainty due to weak lensing. In this analysis, the fiducial model considered is LCDM, as a result, this relation holds.}. The weak lensing uncertainty is modeled using \cite{Hirata2010}:

\begin{equation}
    \frac{\sigma_{WL}}{D_L} = \frac{0.096}{2} \left(\frac{1 - (1 + z)^{-0.62}}{0.62}\right)^{2.36}.
\end{equation}
The total uncertainty in the GW distance measurement is then expressed as:
\begin{equation}
    \sigma_{D_L}^2 = \sigma_{GW}^2 + \sigma_{WL}^2,
\end{equation}
where \( \sigma_{GW} \) reflects detector noise and parameter estimation errors, while \( \sigma_{WL} \) quantifies the lensing contribution. Another potential source of redshift uncertainty arises from peculiar velocities of host galaxies, especially at low redshifts (\( z \lesssim 0.1 \)), where local gravitational interactions perturb the redshift from the pure Hubble flow. However, since the majority of the GW events considered in this study are located at \( z \gtrsim 0.1 \), this effect is subdominant and does not significantly impact the overall error budget \cite{Nimonkar:2023pyt}.

The joint posterior distributions for the parameters \( H_0 \), \( w_0 \), and \( w_a \) are shown in Figure~\ref{fig:BACornerPlot}. The reconstructed evolution of the DE EoS parameter \( (w_0, w_a) \) is presented in Figure~\ref{fig:BAParamEvo}. The uncertainties in \( w_0 \) and \( w_a \) show a characteristic redshift dependence. They are largest at low redshift due to limited sensitivity to the evolving component, reach a minimum near \( z \sim 1 \) where the GW event rate and SNR peak, and increase again at higher redshifts due to the reduced number of detectable events \cite{Zhao:2010sz}. Notably, NSBH systems yield slightly tighter constraints than BNS sources, owing to their higher SNRs from more massive black holes, which enhance the precision of \( D_L\) measurements. The redshift at which EoS parameter uncertainties are minimized also depends on the assumed delay time distribution of GW sources. In this study, we adopt a minimum delay time of 500 Myr. Varying this delay would shift the redshift distribution of events and thereby the optimal sensitivity window for constraining DE dynamics. Also, the assumption of the mass distribution can impact the number of detectable events, and can change the result slightly. However, for the SNR cutoff of 20 considered in this analysis, number of detected events will not change significantly.   

These results demonstrate that next-generation GW detectors such as CE and ET will provide powerful constraints on the time evolution of DE, even for simple two-parameter extensions to the cosmological constant model, through precise bright siren observations. Overall, starting from a fiducial \( \Lambda \)CDM cosmology with parameters consistent with \cite{Planck:2015fie}, our analysis demonstrates that bright siren observations from next-generation GW detectors can constrain the EoS parameters \( w_0 \) and \( w_a \) at sub-percent and percent-level precision, respectively. For the Barboza--Alcaniz parametrization, we find uncertainties of \( \sigma(w_0) \sim 0.002 \) and \( \sigma(w_a) \sim 0.004 \) for NSBH sources, with slightly broader posteriors for BNSs. A summary of all inferred cosmological parameters across the DE models studied is provided in Table~\ref{tab:ModelComparisn} and is  shown in Figure \ref{fig:summary}.

\subsection{Physically Motivated Scalar Field Models}
\label{sec:HilltopResults}

We now present results for the hilltop quintessence model, a physically motivated scalar field scenario that exhibits thawing behavior in the late Universe, see Section~\ref{sec:phymot}. In this analysis, we fix the dimensionless curvature parameter \( K \), which characterizes the steepness of the potential near the hilltop. We then infer the remaining three parameters: the Hubble constant \( H_0 \), the present-day DE equation of state \( w_0 \), and the present-day scalar field density \( \Omega_{\phi0} \). These parameters govern the evolution of the DE equation of state \( w(z) \) in the thawing scenario.
The joint posterior distribution is given by:
\begin{equation}
\begin{aligned}
    P(H_0, w_0, \Omega_{\phi0}) &\propto \Pi(H_0)\Pi(w_0)\Pi(\Omega_{\phi0})\prod_{i=1}^{n_{GW}} \mathcal{L} \left(D_L^{i} \mid H_0, w_0, \Omega_{\phi0}, z^i \right),
\end{aligned}
\label{eq:HilltopPosteriorFixedK}
\end{equation}
where \( K \) is fixed to its fiducial value (chosen here to be \( K = 1.2 \)). This choice corresponds to a moderate thawing regime in quintessence models, consistent with previous studies (e.g., \cite{Dutta:2008px, Caldwell:2005tm}), enabling detectable deviations from \( \Lambda \)CDM while avoiding strong degeneracies. Fixing \( K \) leads to tighter constraints on the remaining parameters but introduces a modeling assumption. As demonstrated in Appendix~\ref{sec:appendix:hilltop4param}, allowing \( K \) to vary broadens the posteriors of \( w_0 \) and \( \Omega_{\phi0} \) due to their coupled evolution with the scalar field. The redshifts \( z^i \) are assumed to be known precisely through spectroscopic EM counterparts and are treated as delta-function posteriors. Luminosity distance uncertainties are incorporated as in Section~\ref{sec:PhenomenologicalResults}.

\begin{figure}[ht]
    \centering
    \includegraphics[width=10.0cm, height=9.0cm]{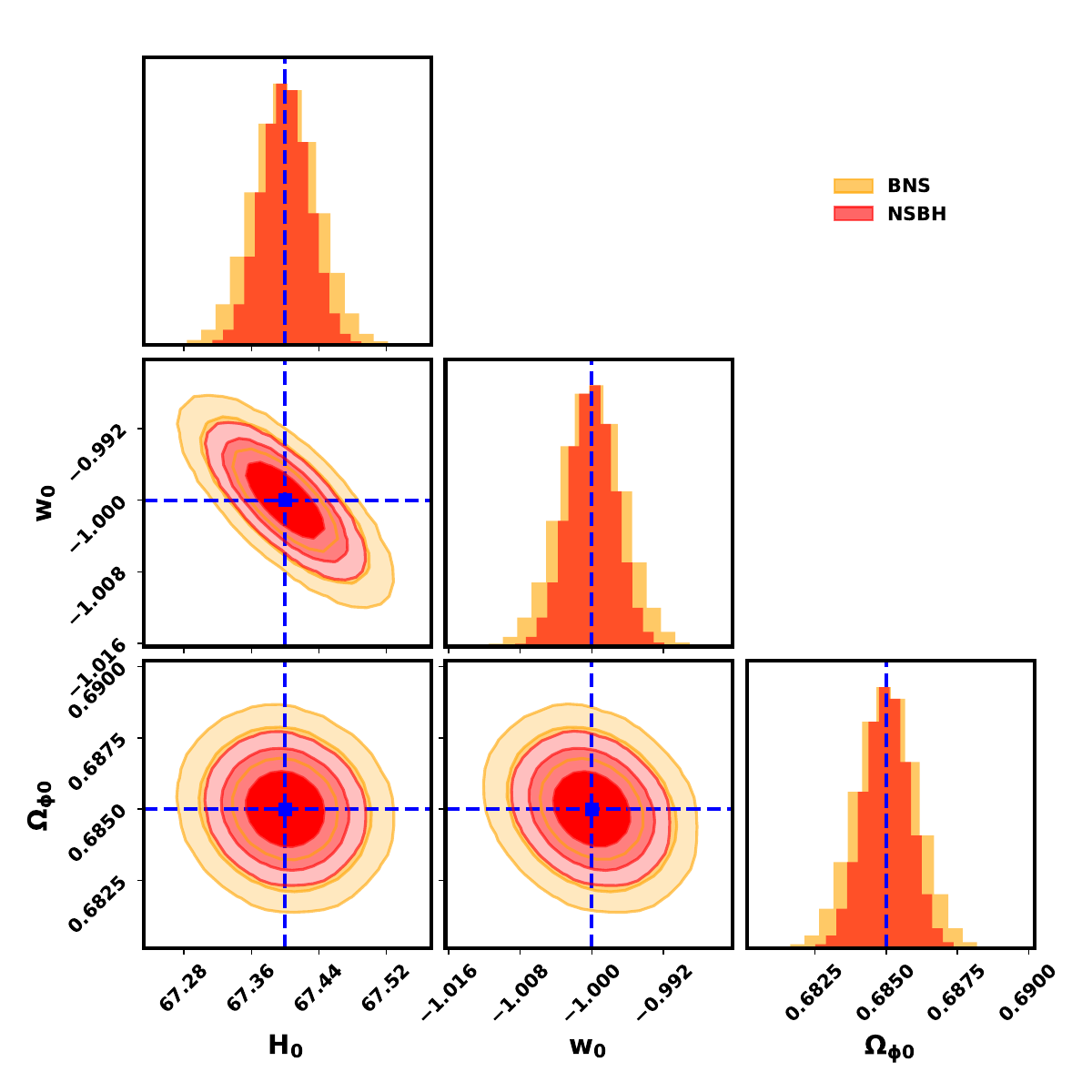}
    \caption{\small Joint posterior distributions of the hilltop quintessence parameters \( w_0 \), \( \Omega_{\phi0} \), and the Hubble constant \( H_0 \), with the value of the parameter \( K \) kept fixed. The contours correspond to the 68.27\% (1$\sigma$), 95.45\% (2$\sigma$), and 99.73\% (3$\sigma$) confidence levels for the two-dimensional marginalized posteriors.}
    \label{fig:HilltopCornerFixedK}
\end{figure}

\vspace{0.5em}
\noindent
Figure~\ref{fig:HilltopCornerFixedK} presents the posterior distributions for \( H_0 \), \( w_0 \), and \( \Omega_{\phi0} \). Compared to the full four-parameter case, the uncertainties are significantly reduced due to the fixed value of \( K \), which otherwise introduces degeneracies, particularly with \( w_0 \). A mild positive correlation is observed between \( \Omega_{\phi0} \) and \( w_0 \), consistent with the scalar field dynamics: larger present-day energy densities require more negative values of \( w_0 \) to yield a similar expansion history. 

\begin{figure}[ht]
    \centering
    \includegraphics[width=11.0cm, height=10.0cm]{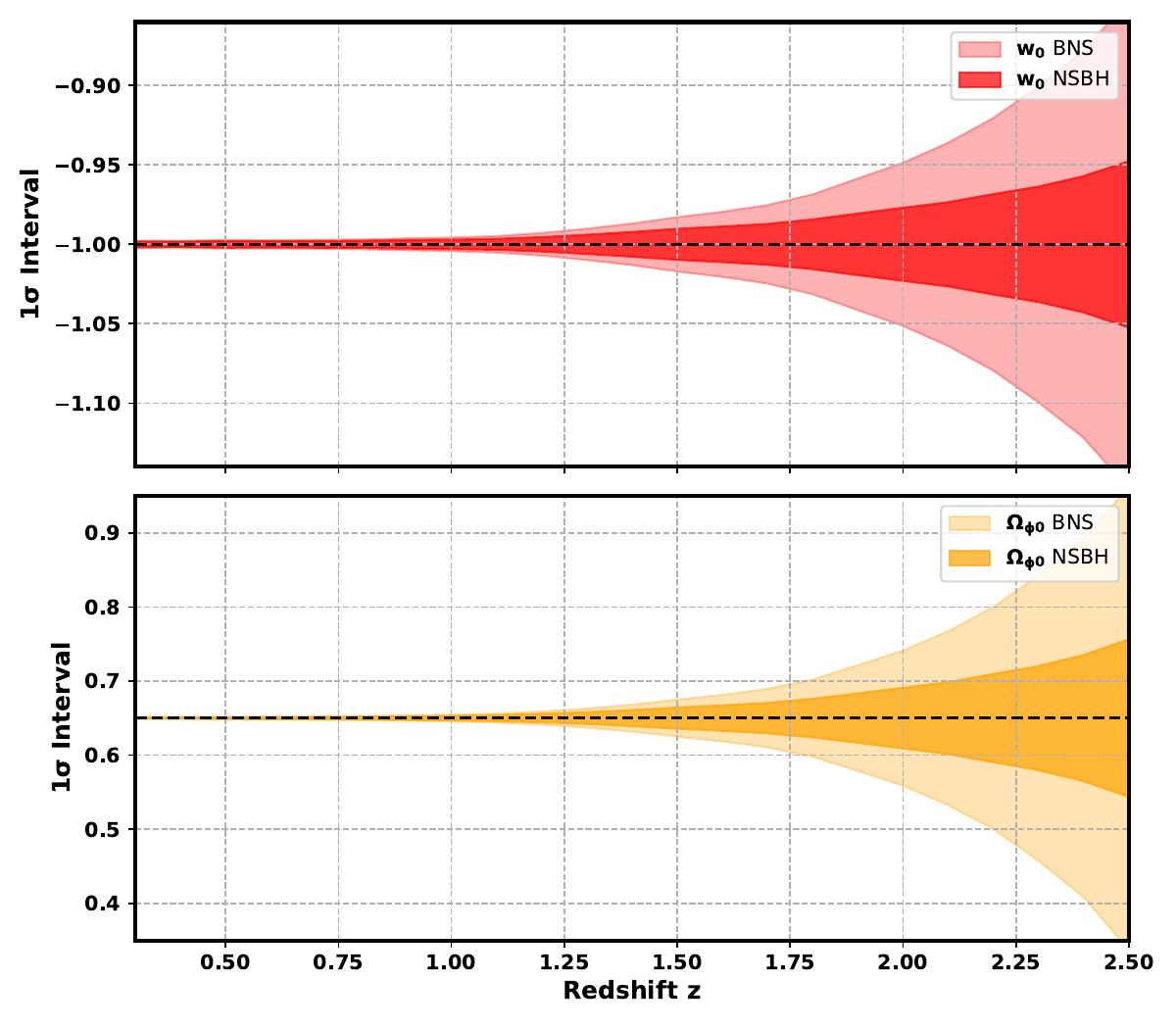}
    \caption{\small Reconstructed redshift evolution of the DE equation of state parameter (\( w_0, \Omega_{\Phi_0} \)) in the hilltop quintessence model with fixed \( K = 1.2 \). The shaded regions represent 1\(\sigma\) uncertainties for BNS and NSBH populations. NSBH systems provide tighter constraints due to their higher SNRs.}
    \label{fig:HilltopEoSEvolutionFixedK}
\end{figure}

\vspace{0.5em}
\noindent
The evolution of the reconstructed (\( w_0, \Omega_{\Phi_0} \)) is shown in Figure~\ref{fig:HilltopEoSEvolutionFixedK}. As expected for thawing models, the shape of this evolution is now primarily driven by \( w_0 \) and \( \Omega_{\phi0} \), while \( K \) determines the overall steepness of the field's potential. Fixing \( K \) leads to a reduction in uncertainty across the redshift range, particularly at higher redshifts where its degeneracy with \( w_0 \) would otherwise broaden the credible interval. The smallest uncertainties occur around \( z \sim 1 \), where the number and SNR of GW events peak. NSBH sources provide more precise constraints than BNS, as expected from their higher mass and associated luminosity distance precision. Overall, by fixing the curvature parameter \( K \), we obtain sub-percent and percent-level constraints on the remaining model parameters. For NSBH sources, we find uncertainties of approximately \( \sigma(w_0) \sim 0.004 \) and \( \sigma(\Omega_{\phi0}) \sim 0.002 \), while for BNS sources the uncertainties are slightly larger. These results highlight the constraining power of next-generation bright siren observations even in the context of physically motivated scalar field models. A full comparison of parameter uncertainties across all models is provided in Table~\ref{tab:ModelComparisn}. A more general analysis where \( K \) is treated as a free parameter is presented in Appendix~\ref{sec:appendix:hilltop4param}, where we analyse  the impact of relaxing this assumption and highlight the degeneracies that arise among the full parameter set \( \{H_0, w_0, K, \Omega_{\phi0}\} \).

\vspace{0.5em}
\noindent

\subsection{Evolving Dark Matter Models}
\label{sec:EvolvingDMResults}

We finally consider an evolving dark matter model in which the DE equation of state \( w(z) \) includes an additional parameter \( w_b \) (as defined in Equation \eqref{eq:ExtCPL}), and the dark matter sector evolves with redshift through a coupling parameter \( \alpha \). This model extends the standard \( w_0 \)-\( w_a \) parametrization by introducing an extra term \( w_b \): it  allows the dark matter equation of state to vary, governed by the parameter \( \alpha \). In this analysis, we jointly infer five parameters: \( H_0 \), \( w_0 \), \( w_a \), \( w_b \), and \( \alpha \). See Section~\ref{sec:evdm}. The corresponding posterior is given by

\begin{figure}[ht]
    \centering
    \includegraphics[width=13.0cm, height=12.0cm]{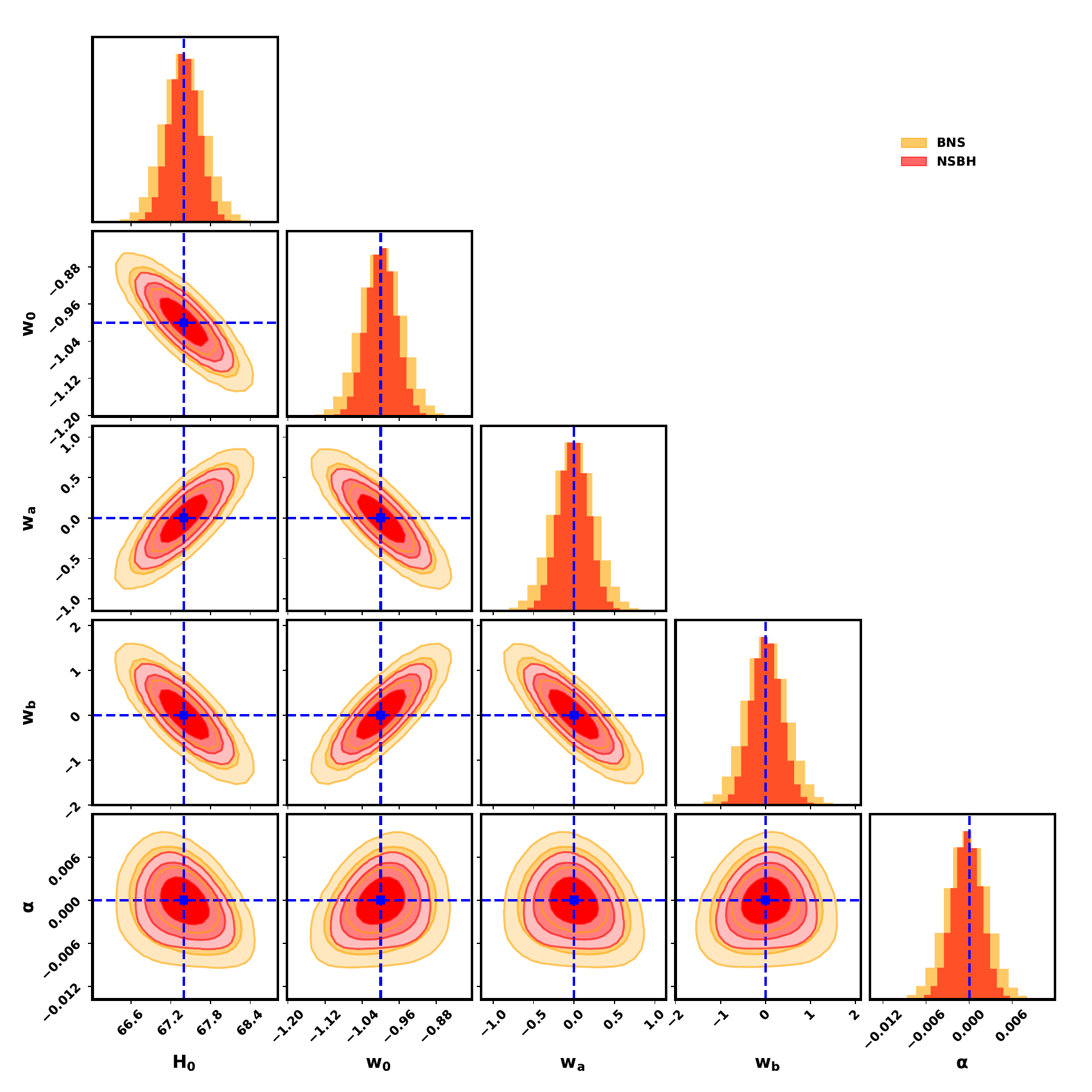}
    \caption{\small Joint posterior distributions for the evolving dark matter model parameters: \( w_0 \), \( w_a \), \( w_b \), \( \alpha \), and \( H_0 \), inferred from simulated bright siren data including BNS and NSBH sources. The contours correspond to the 68.27\% (1$\sigma$), 95.45\% (2$\sigma$), and 99.73\% (3$\sigma$) confidence levels for the two-dimensional marginalized posteriors.}
    \label{fig:DEDMCornerPlot}
\end{figure}

\begin{equation}
\begin{aligned}
    P(H_0, w_0, w_a, w_b, \alpha) &\propto \Pi(H_0) \Pi(w_0) \Pi(w_a) \Pi(w_b) \Pi(\alpha)\prod_{i=1}^{n_{GW}} \mathcal{L} \left(D_L^{i} \mid H_0, w_0, w_a, w_b, \alpha, z^i \right),
\end{aligned}
\label{eq:EDMposterior}
\end{equation}
where the redshift \( z^i \) is obtained from spectroscopic EM counterparts, and hence modeled as a delta function. All other observational assumptions remain consistent with those discussed in Section~\ref{sec:PhenomenologicalResults}.

\begin{figure}[ht]
    \centering
    \includegraphics[width=11.0cm, height=13.0cm]{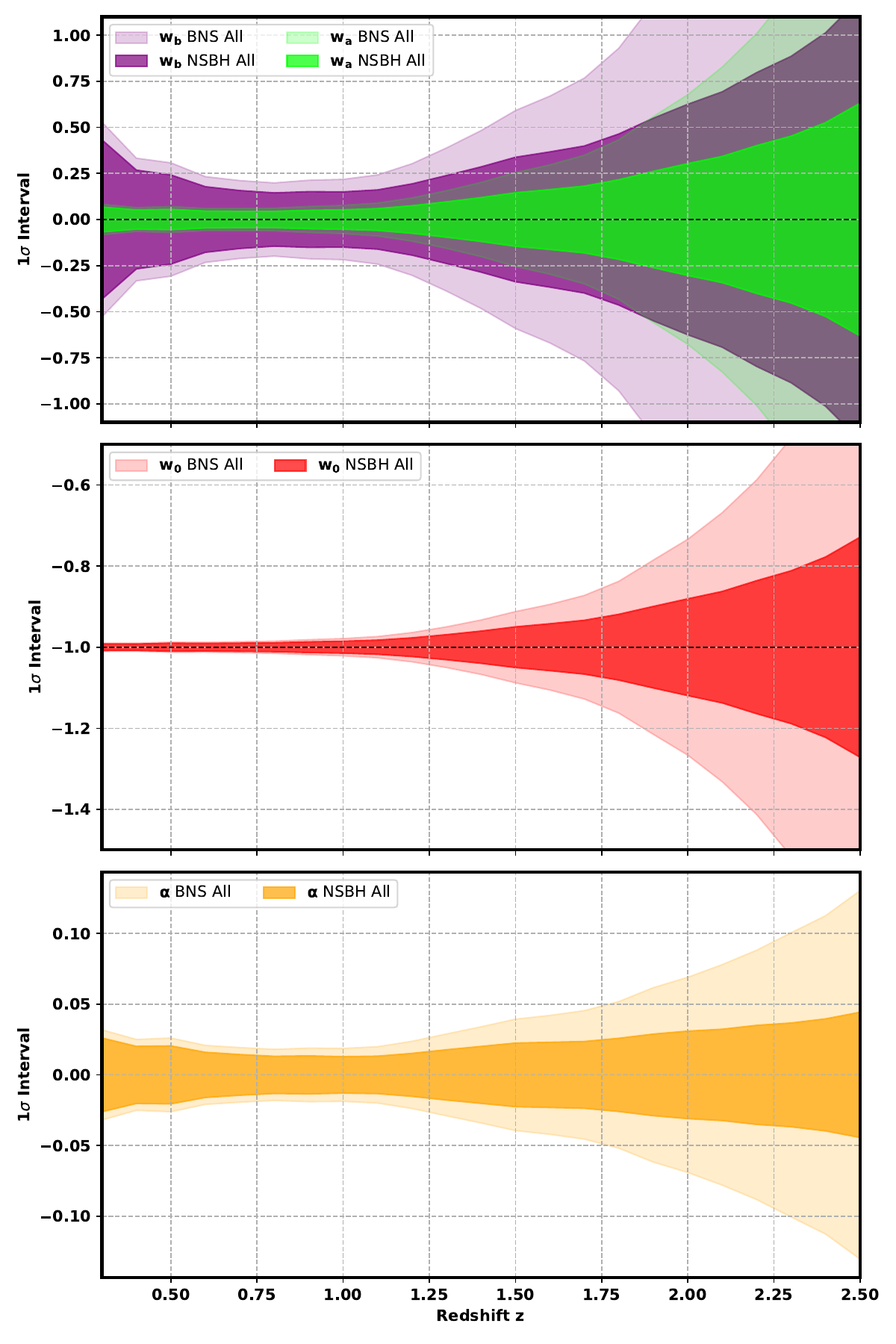}
    \caption{\small Reconstructed evolution of the DE equation of state \( w(z) \) in the evolving dark matter model. Shaded bands represent the 1\(\sigma\) credible intervals for BNS and NSBH sources.}
    \label{fig:DEDMParamEvo}
\end{figure}

Figure~\ref{fig:DEDMCornerPlot} shows the joint posterior distributions for the five model parameters. The corner plot reveals that the parameters \( w_0, w_a \), and \( w_b \) exhibit correlations among each other, while the coupling parameter \( \alpha \) is degenerate  with all the parameters. The redshift evolution of the DE equation of state is shown in Figure~\ref{fig:DEDMParamEvo}, where the 1\(\sigma\) confidence intervals for both BNS and NSBH populations are presented. Notably, the parameter \( w_0 \), which governs the low-redshift behavior of the EoS, is better constrained compared to the high-redshift parameters \( w_a \) and \( w_b \). The evolution of the EoS becomes more prominent at intermediate redshifts (\( z \sim 1 \)) where the GW event distribution peaks.

The uncertainty bands in Figure~\ref{fig:DEDMParamEvo} further highlight that NSBH sources provide tighter constraints across redshift, consistent with their higher SNR and improved distance measurements relative to BNS events. Among the EoS parameters, \( w_b \), which dominates the high-redshift behavior, shows the widest uncertainty, reflecting the lower density of sources at high \( z \). The coupling parameter \( \alpha \) has been marginalized over in the plot but its posterior was found to be symmetric and centered around zero, consistent with minimal deviation from standard non-interacting dark matter. Quantitatively, the evolving dark matter model exhibits broader posterior distributions compared to simpler parametrizations due to the increased number of free parameters and associated degeneracies. Nevertheless, meaningful constraints can still be obtained. For NSBH sources, the uncertainty on \( w_0 \) is at the few percent level (\( \sigma(w_0) \sim 0.03 \)), while \( w_a \), \( w_b \), and \( \alpha \) are constrained to within \( \sim 0.18 \), \( \sim 0.33 \), and \( \sim 0.002 \), respectively. These results demonstrate that even for extended models involving DE-dark matter interactions, bright siren observations from CE and ET can yield informative posteriors. A complete summary of the inferred parameter constraints across all models is provided in Table~\ref{tab:ModelComparisn}. 
These results demonstrate that even extended models involving coupled DE-dark matter dynamics can be effectively constrained using bright siren data from next-generation GW observatories. While we focussed on the specific setup of Section~\ref{sec:evdm}, our methods can be applied to other scenarios as well. The ability to jointly infer the full set of parameters with meaningful posterior bounds, despite the increased model complexity, underscores the strength of hierarchical inference techniques applied to multi-messenger cosmology.

\subsection{Model Comparison}
\label{sec:modelcomparison}

To assess the constraining power and interpretability of different DE models, we compare the results obtained for the three scenarios considered in this work: (i) the phenomenological Barboza-Alcaniz (BA) parametrization discussed in section \ref{sec:thphpar}, (ii) the physically motivated hilltop quintessence model discussed in section \ref{sec:phymot}, and (iii) the evolving dark matter model discussed in section \ref{sec:evdm}. Each framework captures different aspects of DE dynamics and introduces distinct parameterizations with increasing levels of theoretical structure.

\begin{figure}[ht]
    \centering
    \includegraphics[width=15.0cm, height=5.0cm]{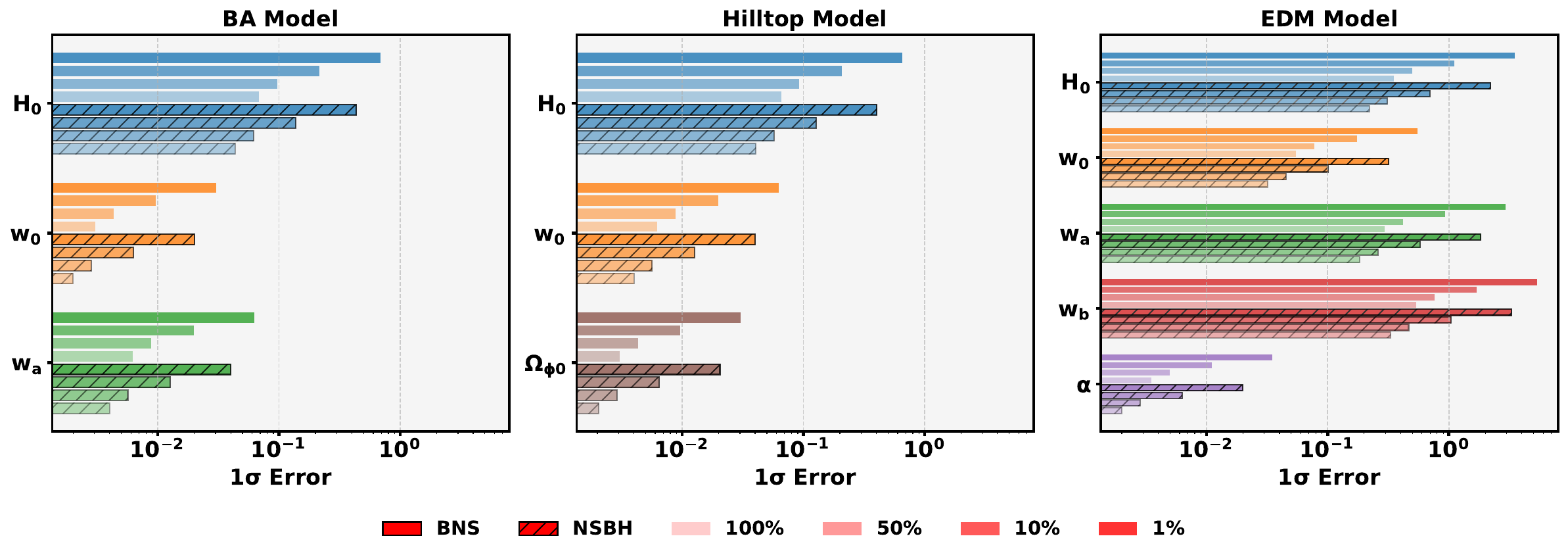}
    \caption{\small 1$\sigma$ uncertainties on dark energy parameters across different models, source types, and EM counterpart detection fractions. Each panel corresponds to a dark energy model Barboza-Alcaniz (left), Hilltop Quintessence (center), and Evolving Dark Matter (right). For each model, we show the marginalized posterior standard deviation for relevant parameters, using binary neutron star (BNS, solid bars) and neutron star–black hole (NSBH, hatched bars) sources. The color shading encodes the fraction of GW sources with identified EM counterparts, ranging from 1\% (lightest) to 100\% (darkest), illustrating how improved EM counterpart detection enhances parameter constraints. The x-axis is logarithmic to emphasize variations across several orders of magnitude. Parameters not included in a model are omitted.}
    \label{fig:summary}
\end{figure}

The BA parametrization provides a flexible, two-parameter extension to the cosmological constant, ensuring regular behavior of the EoS across redshift. It is purely phenomenological and easy to constrain, but lacks direct physical motivation. The hilltop quintessence model, on the other hand, is grounded in scalar field dynamics and introduces a direct connection to fundamental physics via the curvature of the potential and energy density of the scalar field. The evolving dark matter model includes both a phenomenological EoS and a coupling term to evolving dark matter, allowing for a broader class of interactions in the dark sector.

Figure~\ref{fig:summary} provides a visual summary of the parameter estimation uncertainties across all three dark energy models considered in this work. It illustrates the evolution of 1$\sigma$ errors as a function of the EM counterpart detection fraction, for both BNS and NSBH sources. The plot highlights that even a modest fraction of EM-bright events can lead to percent-level precision on key parameters like \( H_0 \) and \( w_0 \), with further improvements as the detection rate increases. Notably, while the Barboza--Alcaniz model yields the tightest constraints overall, the scalar field and evolving dark matter models are also meaningfully constrained despite their larger or more theoretically motivated parameter spaces. This figure reinforces the central result of our analysis: bright sirens observed by next-generation GW detectors will enable high-precision cosmological inference across a wide class of dark energy models, provided that EM follow-up is available for a reasonable fraction of events.

\begin{table*}[ht]
\renewcommand{\arraystretch}{1.5}
\setlength{\tabcolsep}{4pt}
\small  

\centering
\begin{tabular}{|l|c|c|c|c|c|c|}
    \hline
    \multicolumn{7}{|c|}{\textbf{Using BNS as bright sirens}} \\
    \hline
    \textbf{Model} & \( \mathbf{H_0} \) & \( \mathbf{w_0} \) & \( \mathbf{w_a} \) & \( \mathbf{w_b} \) & \( \boldsymbol{\alpha} \) & \( \boldsymbol{\Omega_{\phi0}} \) \\
    \hline
    Barboza\text{-}Alcaniz (BA) & $67.4^{+0.069}_{-0.068}$ & $-1.0^{+0.0031}_{-0.0030}$ & $0.0^{+0.0062}_{-0.0063}$ & -- & -- & -- \\
    \hline
    Hilltop Quintessence & $67.4^{+0.065}_{-0.066}$ & $-1.0^{+0.0062}_{-0.0063}$ & -- & -- & -- & $0.685^{+0.0030}_{-0.0031}$ \\
    \hline
    Evolving Dark Matter &  $67.4^{+0.350}_{-0.354}$ & $-1.02^{+0.056}_{-0.054}$ & $0.001^{+0.290}_{-0.301}$ & $0.013^{+0.531}_{-0.542}$ & $0.0^{+0.0034}_{-0.0036}$ & -- \\
    \hline
\end{tabular}

\vspace{0.15cm}

\begin{tabular}{|l|c|c|c|c|c|c|}
    \hline
    \multicolumn{7}{|c|}{\textbf{Using NSBH as bright sirens}} \\
    \hline
    \textbf{Model} & \( \mathbf{H_0} \) & \( \mathbf{w_0} \) & \( \mathbf{w_a} \) & \( \mathbf{w_b} \) & \( \boldsymbol{\alpha} \) & \( \boldsymbol{\Omega_{\phi0}} \) \\
    \hline
    Barboza\text{-}Alcaniz (BA) & $67.4^{+0.043}_{-0.044}$ & $-1.0^{+0.002}_{-0.002}$ & $0.0^{+0.004}_{-0.004}$ & -- & -- & -- \\
    \hline
    Hilltop Quintessence & $67.4^{+0.041}_{-0.040}$ & $-1.0^{+0.004}_{-0.004}$ & -- & -- & -- & $0.685^{+0.0020}_{-0.0021}$ \\
    \hline
    Evolving Dark Matter &  $67.4^{+0.221}_{-0.220}$ & $-0.99^{+0.032}_{-0.032}$ & $0.001^{+0.183}_{-0.184}$ & $0.014^{+0.332}_{-0.327}$ & $0.0^{+0.0020}_{-0.0020}$ & -- \\
    \hline
\end{tabular}
\caption{\small Inferred values of cosmological parameters for the three DE models studied. All results are based on hierarchical inference using simulated bright siren data.}
\label{tab:ModelComparisn}
\end{table*}

We summarize the inferred parameter constraints for each model in Table~\ref{tab:ModelComparisn}. In each case, the Hubble constant \( H_0 \) is jointly inferred, along with the model-specific parameters. The phenomenological and evolving dark matter models share the parameters \( w_0 \) and \( w_a \), while the scalar field model uses \( w_0 \) in a more constrained theoretical context. The evolving dark matter model includes two additional degrees of freedom, \( w_b \) and the coupling parameter \( \alpha \), while the scalar field model includes \( K \) and \( \Omega_{\phi0} \), which control the shape and initial condition of the potential. From a statistical perspective, all three models yield consistent estimates of the Hubble constant, with uncertainties dominated by the precision of luminosity distance measurements. The BA model achieves tight constraints on \( w_0 \) and \( w_a \), particularly at intermediate redshifts where the GW event density peaks. The scalar field model produces slightly broader posteriors due to its more constrained functional form, though it offers a clear connection to fundamental theory via \( K \) and \( \Omega_{\phi0} \). The evolving dark matter model, despite its larger parameter space, is still able to constrain the full set of five parameters meaningfully, demonstrating the power of GW bright siren data in exploring extended dark sector physics.

Overall, our analysis highlights the versatility of GW cosmology in testing a wide range of DE scenarios, from phenomenological parametrizations to physically motivated scalar fields and non-trivial dark matter evolution. Future observations will further sharpen these constraints and allow for robust model selection in the DE sector by means of methods
completely independent from other cosmological probes.

\section{Conclusions}
\label{sec:conclusion}

In this work, we investigated the capability of next-generation gravitational wave (GW) observatories to constrain the evolution of the DE equation of state using bright standard sirens-binary neutron star (BNS) and neutron star–black hole (NSBH) mergers with electromagnetic counterparts. By focusing exclusively on bright sirens, we leveraged the spectroscopically determined redshifts of host galaxies to achieve precise and accurate measurements of the luminosity distance–redshift relation without requiring a cosmic distance ladder. We considered three representative  classes of DE models: (1) a phenomenological parametrization based on the Barboza--Alcaniz extension, (2) a physically motivated scalar field model inspired by hilltop quintessence, and (3) an evolving dark matter scenario in which the DE equation of state couples dynamically to the dark matter sector. Using a hierarchical Bayesian framework, we jointly inferred the Hubble constant \( H_0 \) and the corresponding DE parameters specific to each model adopting realistic observational uncertainties, including both detector noise and astrophysical effects such as weak lensing.

For each model, we reconstructed the redshift evolution of the DE equation of state \( w(z) \) and examined how it responds to the underlying model parameters. The phenomenological approach offered tight constraints on \( w_0 \) and \( w_a \) across redshifts, while the hilltop quintessence model enabled a physically grounded reconstruction of thawing scalar field dynamics through the parameters \( K \) and \( \Omega_{\phi0} \). The evolving dark matter model, with its extended parameter space, demonstrated that even more complex dark sector interactions can be probed effectively using bright siren data alone. Our results show that next-generation GW detectors such as the Einstein Telescope (ET) and Cosmic Explorer (CE), combined with precise spectroscopic redshift measurements, will enable high-precision cosmological inference. The redshift evolution of DE can be mapped with percent-level accuracy, allowing us to distinguish between different theoretical models, including those motivated by scalar field dynamics and dark sector interactions.

Unlike the previous analysis \cite{Afroz:2024joi}, which explored both bright and dark sirens using combined data from LSST and DESI, this study focuses solely on the bright siren channel and adopts a diverse suite of DE parameterizations. This targeted approach demonstrates that even without galaxy cross-correlation or dark siren statistics, bright sirens alone can place competitive constraints on DE models - provided that high-quality EM counterparts and spectroscopic redshifts are available. Looking ahead, the synergy between gravitational wave observations and electromagnetic follow-up will continue to be a cornerstone of multi-messenger cosmology. Our work highlights the critical role of bright sirens as cosmological probes and emphasizes the importance of spectroscopic surveys in maximizing the scientific return from GW detections. As future facilities come online and observation time increases, we expect the precision of DE measurements to improve further, opening new windows into the physics of cosmic acceleration and the fundamental nature of the universe.

\section*{Acknowledgements}
It is a pleasure to thank Michele Mancarella and Ivonne Zavala for useful discussions. 
The work of SA and SM is part of the \texttt{⟨data|theory⟩ Universe-Lab}, supported by TIFR and the Department of Atomic Energy, Government of India.  GT is partially funded by the STFC grants ST/T000813/1 and ST/X000648/1. The authors express gratitude to the computer cluster of \texttt{⟨data|theory⟩ Universe-Lab} for computing resources. The authors are thankful to the Cosmic Explorer and Einstein Telescope collaboration for providing the noise specifications. Part of this work was carried out during the 2025 {The Dawn of Gravitational Wave Cosmology} workshop, supported by the Fundaci\'on Ramon Areces and hosted by the {Centro de Ciencias de Benasque Pedro Pascual}. We thank both the CCBPP and the Fundaci\'on Areces for providing a stimulating and highly productive research environment. We acknowledge the use of the following packages: Astropy \cite{robitaille2013astropy, price2018astropy}, Bilby \cite{ashton2019bilby}, Pandas \cite{mckinney2011pandas}, NumPy \cite{harris2020array}, Scipy \cite{virtanen2020scipy}, Dynesty \cite{speagle2020dynesty}, emcee \cite{foreman2013emcee} and Matplotlib \cite{Hunter:2007}

\appendix
\section{Extended Results for Hilltop Quintessence: Full 4-Parameter Inference}
\label{sec:appendix:hilltop4param}

In this appendix, we present the full inference analysis for the hilltop quintessence model in which all four model parameters are allowed to vary: the Hubble constant \( H_0 \), the present-day value of the DE equation of state \( w_0 \), the potential curvature parameter \( K \), and the present-day scalar field energy density \( \Omega_{\phi0} \). These parameters govern the thawing behavior of the scalar field and the resulting dynamics of the DE equation of state.

\begin{figure}[ht]
    \centering
    \includegraphics[width=12.0cm, height=11.0cm]{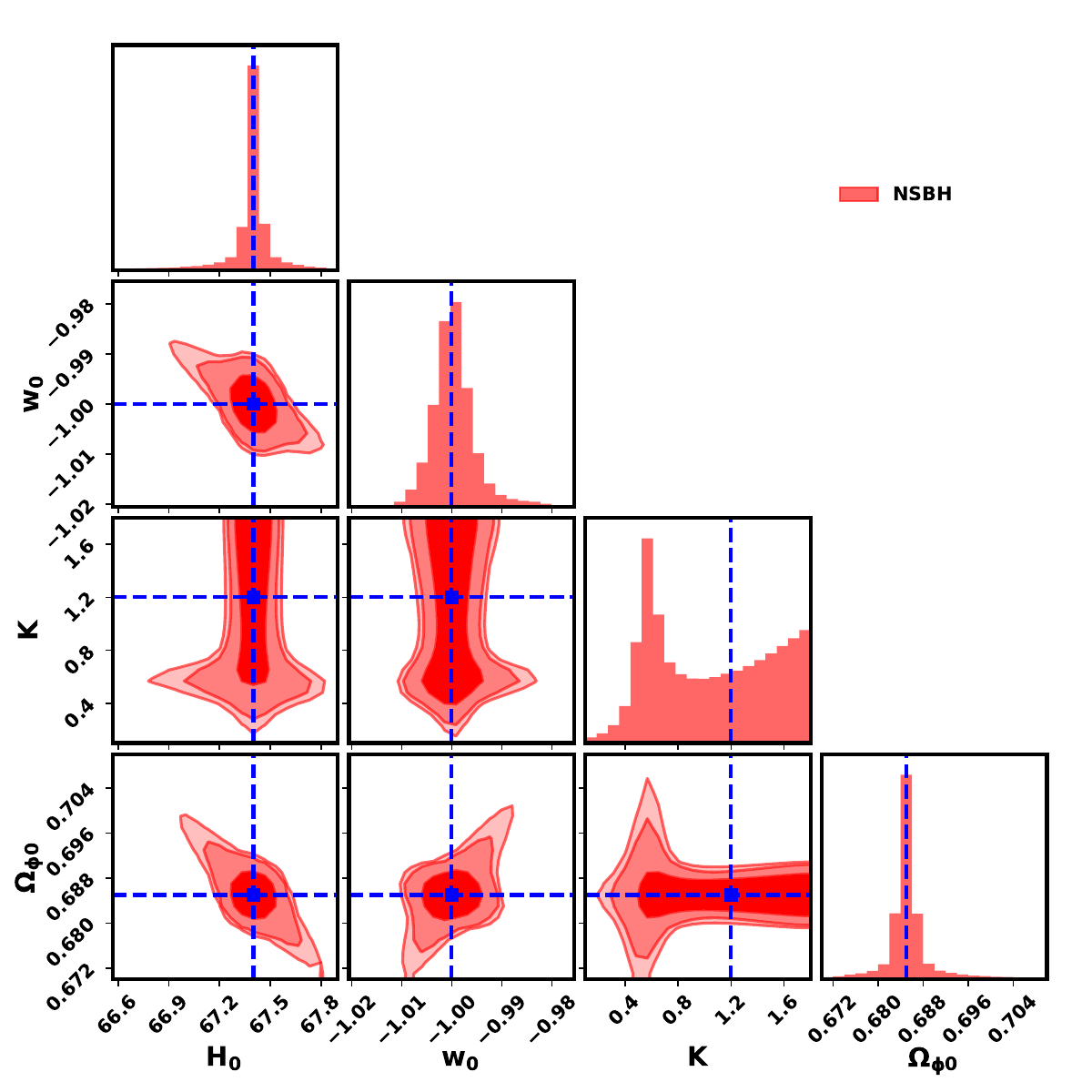}
    \caption{\small Joint posterior distributions of the hilltop quintessence parameters \( H_0 \), \( w_0 \), \( K \), and \( \Omega_{\phi0} \), inferred from simulated bright siren NSBH observations. The contours correspond to the 68.27\% (1$\sigma$), 95.45\% (2$\sigma$), and 99.73\% (3$\sigma$) confidence levels for the two-dimensional marginalized posteriors.}
    \label{fig:HilltopCornerNSBH}
\end{figure}

The joint posterior distribution is given by:
\begin{equation}
\begin{aligned}
    P(H_0, w_0, K, \Omega_{\phi0}) &\propto \Pi(H_0)\Pi(w_0)\Pi(K)\Pi(\Omega_{\phi0}) \prod_{i=1}^{n_{GW}} \mathcal{L} \left(D_L^{i} \mid H_0, w_0, K, \Omega_{\phi0}, z^i \right),
\end{aligned}
\label{eq:HilltopPosterior}
\end{equation}
where \( z^i \) denotes the spectroscopic redshifts of the GW sources, treated as delta-function posteriors. The observational setup-including treatment of luminosity distance uncertainties and event selection-is consistent with the discussion in Section~\ref{sec:PhenomenologicalResults}.

\begin{figure}[ht]
    \centering
    \includegraphics[width=12.0cm, height=11.0cm]{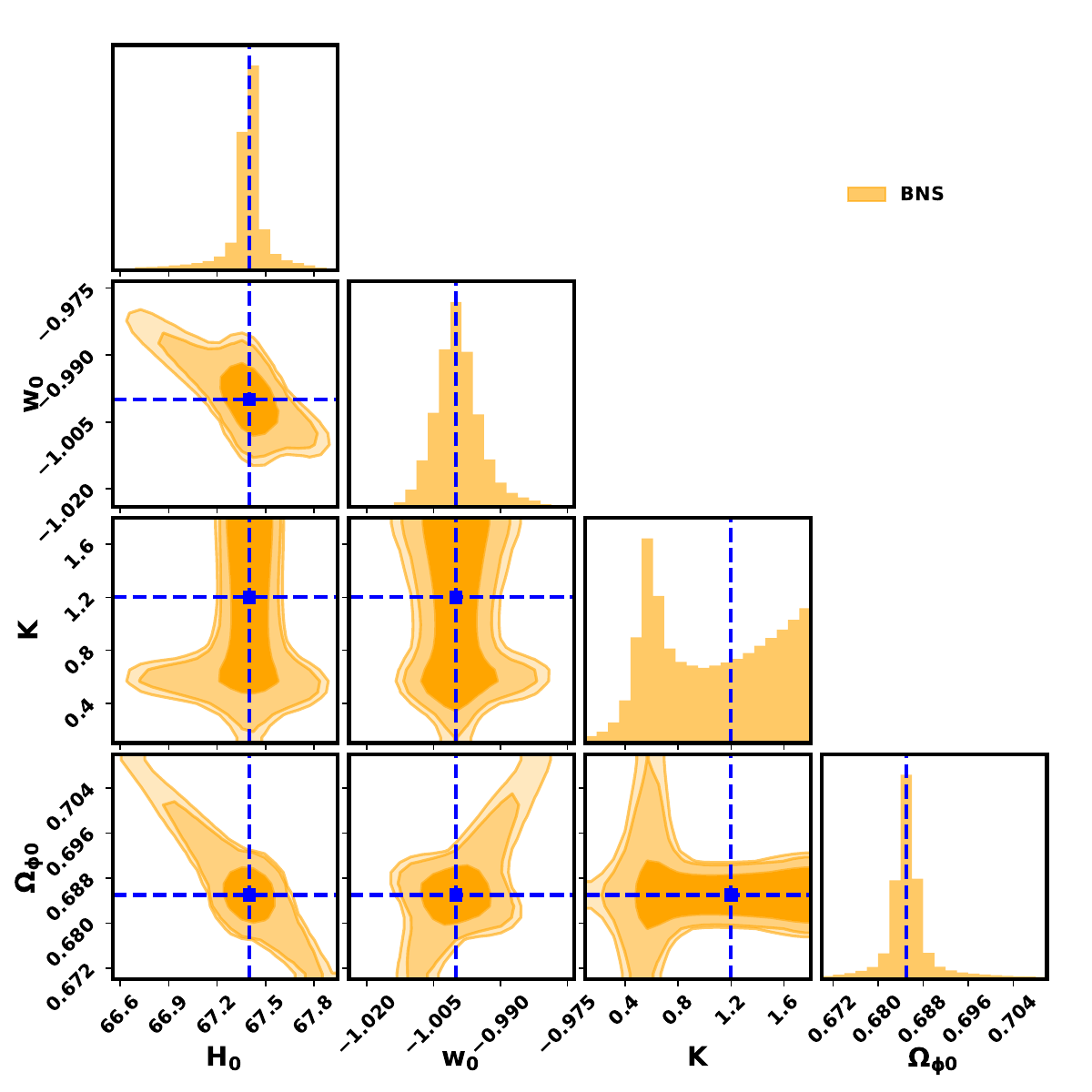}
    \caption{\small Joint posterior distributions of the hilltop quintessence parameters \( H_0 \), \( w_0 \), \( K \), and \( \Omega_{\phi0} \), inferred from simulated bright siren BNS observations. The contours correspond to the 68.27\% (1$\sigma$), 95.45\% (2$\sigma$), and 99.73\% (3$\sigma$) confidence levels for the two-dimensional marginalized posteriors.}
    \label{fig:HilltopCornerBNS}
\end{figure}

\vspace{0.5em}
\noindent

Figures~\ref{fig:HilltopCornerNSBH} and~\ref{fig:HilltopCornerBNS} display the inferred joint posterior distributions of the hilltop quintessence parameters for NSBH and BNS sources, respectively. As anticipated from the scalar field dynamics, characteristic degeneracies are observed. A notable correlation exists between the present-day EoS parameter \( w_0 \) and the curvature parameter \( K \): flatter potentials (corresponding to smaller \( K \)) result in slower thawing of the scalar field, which allows \( w(z) \) to remain close to \( -1 \) over cosmic time, thus broadening the allowed range of present-day EoS values. A positive correlation is also evident between \( w_0 \) and \( \Omega_{\phi0} \), as larger scalar field energy densities at late times require more negative \( w_0 \) to yield expansion histories consistent with the data.

\begin{figure}[ht]
    \centering
    \includegraphics[width=11.0cm, height=13.0cm]{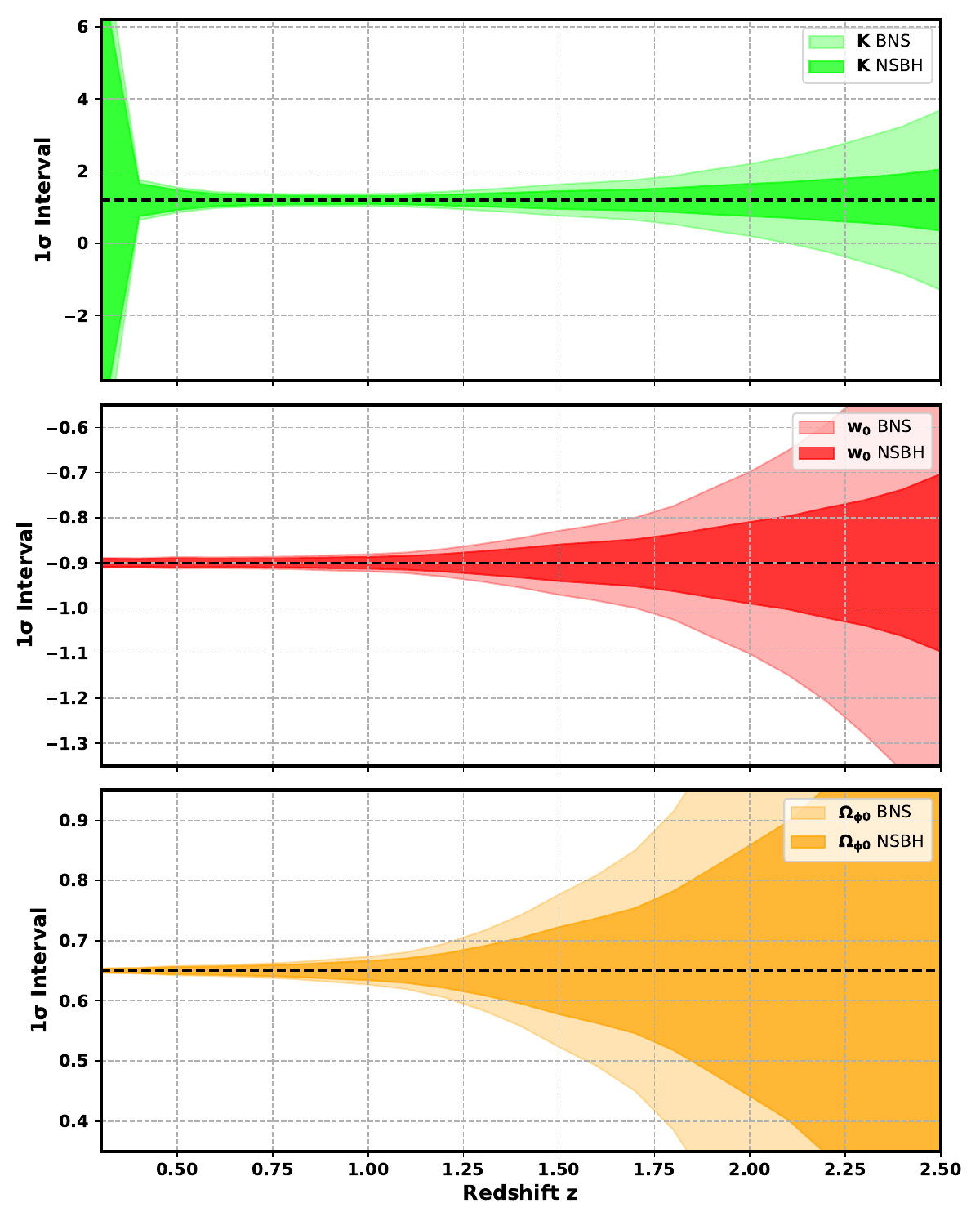}
    \caption{\small Reconstructed redshift evolution of the DE equation of state \( w(z) \) in the hilltop quintessence model using fiducial value \( w_0 = -0.9 \). The shaded bands represent the 68\% confidence intervals for BNS and NSBH sources. As \( w_0 \to -1 \), the evolution becomes indistinguishable from \( \Lambda \)CDM and constraints on \( K \) deteriorate.}
    \label{fig:HilltopEoSEvolution}
\end{figure}

The posterior distribution of the curvature parameter \( K \) displays a rich structure shaped by the underlying scalar field dynamics, with a significantly non-Gaussian posterior. It falls off steeply for small values (\( K \lesssim 0.4 \)), where the shallow curvature near the hilltop causes the scalar field to thaw too early, resulting in significant deviations from \( w(z) \approx -1 \) even at high redshifts. Such rapid evolution is disfavored by the mock data, which prefers expansion histories closer to \(\Lambda\)CDM. The distribution peaks around \( K \sim 0.7 \), where the scalar field exhibits moderate thawing. In this regime, the field remains near the hilltop for most of cosmic history, allowing for mild evolution in \( w(z) \) that still yields Hubble parameter values consistent with the simulated observations.

A mild suppression appears near \( K \sim 1.0 \), likely because the field evolution becomes more pronounced, leading to stronger deviations from the observed expansion history. At larger values (\( K \gtrsim 1.4 \)), the scalar field becomes effectively frozen near the top of the potential, asymptotically approaching \( w(z) \to -1 \). In this regime, the field mimics a cosmological constant, and the posterior flattens as the data becomes insensitive to variations in \( K \), rendering it effectively unconstrained.

These features reflect the characteristic degeneracies between \( K \), \( w_0 \), and \( \Omega_{\phi0} \), which are intrinsic to hilltop thawing quintessence models. In particular, flatter potentials (smaller \( K \)) permit slower scalar field evolution, thereby broadening the allowed range of present-day equation-of-state values \( w_0 \). A positive correlation also emerges between \( w_0 \) and \( \Omega_{\phi0} \), since larger late-time scalar field energy densities require more negative \( w_0 \) to match the expansion history.

Overall, the observed degeneracy structure underscores the sensitivity of the hilltop model to the scalar field dynamics and illustrates how GW data constrain not just the present-day dark energy parameters, but also the underlying shape of the potential driving cosmic acceleration. To illustrate this, we reconstruct the redshift evolution of the DE EoS \( w(z) \) using the fiducial value \( w_0 = -0.9 \). Fixing \( w_0 \) to a value measurably different from \( -1 \) allows for meaningful constraints on \( K \) and helps visualize the dynamics of thawing behavior.

\vspace{0.5em}
\noindent
As seen in Figure~\ref{fig:HilltopEoSEvolution}, both BNS and NSBH sources allow for the reconstruction of \( w(z) \) with uncertainties minimized around \( z \sim 1 \), where the number of detections and SNRs are highest. NSBH systems provide tighter constraints overall, primarily due to higher mass and thus more precise distance measurements. The evolution of \( w(z) \) captures the thawing behavior of the field, gradually deviating from \( -1 \) as redshift decreases. However, as \( w_0 \) approaches the cosmological constant value \( -1 \), the field becomes observationally indistinct from \( \Lambda \)CDM, and \( K \) becomes unconstrained.
This analysis underscores the challenges of constraining thawing quintessence models in the regime close to \( w_0 = -1 \), where degeneracies with \( K \) dominate. When \( w_0 \) is allowed to vary freely, it is necessary to account for these degeneracies. Fixing \( K \), as we do in the main text, leads to significantly tighter bounds on \( w_0 \) and \( \Omega_{\phi0} \), and isolates the role of the scalar field energy density in shaping the DE evolution.

\bibliographystyle{unsrt}
\bibliography{references}
\end{document}